\documentclass[epj,final]{svjour}
\usepackage{epsfig}
\usepackage{amsmath,amssymb}
\begin{document} 
\title{Interaction versus dimerization in one-dimensional Fermi systems}
\dedication{Dedicated to J. Zittartz on the
occasion of his 60th birthday}
\author{Cosima Schuster \and Ulrich Eckern}
 \institute{Institut f\"ur Physik, Universit\"at Augsburg, 
  D-86135 Augsburg, Germany}
\abstract{
In order to study  the effect of interaction and 
lattice distortion on quantum coherence in one-dimensional Fermi systems,
we calculate   the ground state energy and the phase sensitivity  
of a ring of interacting spinless fermions  
on a dimerized lattice. Our numerical DMRG studies, in which we keep up to 1000
states for  systems of about 100 sites, are supplemented by analytical considerations 
using bosonization techniques.  
We find a delocalized phase for an attractive interaction, which differs 
from that obtained for  random lattice distortions. The 
extension of this delocalized phase depends strongly on the dimerization
induced modification of the interaction.  Taking into account the harmonic 
lattice 
energy, we find a dimerized ground state for a repulsive interaction only. 
The dimerization is suppressed at half filling,  when the correlation gap 
becomes large.}
\maketitle
\section{Introduction}
Recent experiments 
on  CuGeO$_3$ \cite{Hase93} and NaV$_2$O$_5$ \cite{Isobe96,Weiden97}, 
which show at low temperature a transition to a non-magnetic 
ground state accompanied by a 
structural transition, led to renewed interest in spin-Peierls systems,
as well as in the more general question of structure versus correlation 
induced metal-insulator transitions.
The model of spinless fermions, which we consider here in detail, 
describes certain aspects of both, and it contains, as a special point
in parameter space, the experimentally relevant 
isotropic Heisenberg antiferromagnet. In addition, the model
contains, in the limit of an Ising type interaction, a szenario which
is closer to that found for a more realistic model, including the 
electron spin, namely the Hubbard-Peierls model.
Here we concentrate
on the spinless fermion model, which nevertheless shows a surprisingly 
rich phase diagram (at zero temperature) as a function of interaction
and dimerization.

In the next section, we introduce
the various representations of the model. In Sect.\ 3,
we quantitatively determine the extension of the delocalized 
phase, which was predicted for an intermediate attractive interaction.
There\-by we identify the region where the fermion-phonon coupling 
is irrelevant, i.e.\ the ground state remains extended. In Sect.\ 4, we  
consider a repulsive interaction, where we find that
a stable dimerization develops. For a very strong interaction
the dimerization is  reduced again,
because conflicting ordering occurs. In the summary, Sect.\ 5, we 
present the $u$-$V$ phase diagram, as well as
the results for the dimerized state. 

\section{The model}
As a starting point for the study of a general spin-Peierls system 
in one 
dimension, we consider an  XXZ model with a dimerized interaction 
in a local magnetic field; the latter can be considered as a staggered
magnetic field resulting  from the surrounding chains \cite{Schulz96},
or resulting from random magnetic impurities. In the following,  
 we concentrate 
on the dimerization-induced modification of the interaction, and, for comparison, include 
the magnetic field only occasionally. Furthermore,
we neglect frustration effects which arise from next-nearest-neighbor
couplings.  Thus we start with the following Hamiltonian:
\begin{eqnarray} 
\label{spins} 
H_{\rm spin}&=& -\sum^{M}_{n=1} J_n (u) \left( \sigma^{x}_{n} \sigma^{x}_{n+1} + \sigma^{y}_{n} \sigma^{y}_{n+1}  \right.
  \notag \\    && \left. + \Delta \sigma^{z}_{n} \sigma^{z}_{n+1}\right)
  + \sum^{M}_{n=1} \left(-h_n\sigma^z_n +\frac{K_0}{2}u_n^2 \right),
\end{eqnarray} 
where $J_n (u) =J (1-(-)^n \lambda_J u)$; the local lattice distortion,
$u_n$, is assumed to be given by $u_n=2(x_n-x_n^0)/a=(-1)^nu$.
For the ``clean" XXZ model, i.e.\ for $u=0$, $h_n =0$, one finds three
phases:
a ferromagnetic phase for $\Delta \geq1$, 
a gapless phase for $-1\leq\Delta < 1$, whose low lying excitations 
are given by those of a Luttinger liquid, and an antiferromagnetic phase for 
$\Delta < -1$.
The spin model is experimentally most relevant for
$\Delta = -1$, i.e. the isotropic antiferromagnetic Heisenberg case.
The corresponding fermion model is obtained 
via the Jordan Wigner transformation:
\begin{eqnarray}
\label{Jordan}
  \sigma^{-}_m &=& e^{-{\rm i}\pi\sum^{m-1}_{l=1} n_l} c_m, \qquad \sigma^{+}_m \;=\;
 (\sigma^{-}_m)^{+}, \notag \\
  \sigma^{z}_m &=& 2 n_m -1, \;\; n_m = c^{+}_m c^{}_m \; . 
\end{eqnarray} 
It is customary to change the notation: 
$J=t$, $J\Delta = -V/2$, and $\epsilon_m=-2h_m$; furthermore we now assume
that the coupling to (static) phonons can be varied independently in  
the hopping and the interaction. The result is the following:
\begin{eqnarray}
\label{ferm}
 H_{\rm fermion} &=&  - \sum_{m} t_m (u)
\left( c^+_mc^{\hphantom +}_{m+1} +c^+_{m+1}c^{\hphantom +}_{m} \right)\notag \\ 
     & &+\; \sum_m V_m (u)   n_m n_{m+1}
\notag 
\\ &&+
\sum_m \epsilon_m n_m  +M\frac{K_0}{2}u^2 ,
\end{eqnarray}
where $t_m (u) = t (1-(-)^m\lambda_t u)$, $V_m (u) = V (1-(-)^m\lambda_V u)$.
Furthermore, we consider twisted boundary conditions, 
$c_0 = {\rm e}^{{\rm i} \phi }c_M$.
The length of the chain is $L=Ma$, the number of electrons
$N$, and the filling $n_0=N/M$. In addition, we restrict ourselves to
half filling, and set $t=1$ and $\lambda_t=1$ in some of the formulas below. 
The corresponding instabilities in the fermion model are at $V=-2$,
towards a phase separated state, and 
at $V=2$, to a charge density wave (CDW) state. (Note that at $V=2$, 
the 4$k_F$-backscattering process becomes relevant \cite{Haldane80}.)
Numerically this CDW state is difficult to determine,
because, for finite systems,  the hopping
lifts the twofold  degeneracy of the ground state via a symmetric superposition,
and a uniform density is found.
For the analytic considerations, we use the fact that the system  
is a Luttinger liquid in the gapless phase,
and that the dimerization can be considered a perturbation.
In the bosonized form \cite{Haldane81}, the Hamiltonian can be written 
as follows:
\begin{eqnarray}
\label{boson}
H_{\rm boson} &=& \displaystyle{\int} \frac{{\rm d}x}{2\pi}\ \left\{\frac{v}{g}
\left[\partial_x\varphi(x,t)\right]^2 +
\frac{1}{vg}\left[\partial_t\varphi(x,t)\right]^2 \right.\notag \\
& + & \left.
\frac{2\pi\lambda u}{l_0}\sin{[2\varphi(x,t)]} 
-  \frac{\pi Va}{l_0^2}\cos{[4\varphi(x,t)]} \right\},
\end{eqnarray}
where $\lambda = \lambda_t t +\lambda_V Vn_0/2$, and $l_0$ is a short distance cut-off, $\approx 2a$. Only the first order
correction to the 2$k_F$-process, $\propto Vu$, has been included. 
The Fermi velocity is given by 
$v={[\pi t\sin(2\eta)]/( \pi -2\eta)}$, and
$g=\pi / 4\eta$,
where $\eta$ parameterizes the interaction according to
$V\!=-2t\cos(2\eta)$.  
 Furthermore, 
the density is given by
\begin{equation} 
\label{density}
\frac{1}{2} \sigma^z_{n=x/a}=\rho(x)-n_0=\frac{\partial_x \varphi}{\pi }+
\frac{2}{l_0} \cos{(2k_Fx+2\varphi)}.
\end{equation}
Clearly, the term 
$\propto \cos{[4\varphi(x,t)]}$, describing $4 k_F$ scattering processes, 
is  responsible for the CDW ordering.  
For completeness we like to state that some properties of the dimerized Heisenberg chain, 
 especially the spectrum, can also be obtained  
\cite{Uhrig96} from the
equivalent relativistic model, the massive Thirring model. 

The numerical results, presented below, 
are obtained on the basis of the XXZ model, Eq. \eqref{spins}.
\section{Phase transition at attractive interaction}
The first conjecture that 
the dimerized XXZ model undergoes a transition from  the dimerized spin 
singlet phase to a 
phase with free spins at $\Delta=\sqrt{2}/2$, 
was given  in \cite{Kohmoto81} based on a mapping onto the 
Ashkin-Teller-model. 
There, an estimate of the phase boundary, based on a series analysis, has been given.
(Note the different sign convention.)
For the XXZ model in a staggered magnetic field,
Alcaraz \cite{Alc94} again found this critical point on the basis of predictions of  
conformal field theory. 
Thus it seems that the periodic distortion  
of the $\sigma^z\sigma^z$-interaction does not change the point of the transition. 
In the case of the bosonic model, assuming an interaction of the form
$\sum U_n\cos{n\varphi}$, 
one finds the following renormalization group  equations \cite{Safi97}:
\begin{equation}\label{RGallg} \frac{{\rm d} U_n}{{\rm d} l}=  
U_n\left(2-\frac{n^2}{4}g\right)+\frac{1}{2}\sum_{n_1+n_2=n}U_{n_1}U_{n_2}.\end{equation} 
The critical value or a model containing only a single periodic term, $g=8/n^2$, 
is in accordance with the 
 exact solution of the sine-Gordon model 
\cite{Dashen75,Sklyanin79}. One concludes that for the 
``clean"
case ($u=0$), there exists an instability for $g=1/2$, i.e. $V=2$. 
Consequently, considering the $\sin(2\varphi)$-contribution in \eqref{boson},
we expect interesting effects for $g=2$, i.e. $V=-\sqrt 2$, provided the
dimerization, $u$, assumes a finite value.

Applied to our model,
the critical behavior near the transition between a 
localized and a delocalized state 
is determined by the Berezinski\u i-Kosterlitz-Thouless equations
\cite{Zang95}:
\begin{eqnarray}
\frac{{\rm d} u}{{\rm d} l} = u(2-g) \label{KTa} \\
\frac{{\rm d} g}{{\rm d} l} = -\frac{1}{2}g^2\lambda^2u^2.\label{KTb}
\end{eqnarray}
Considering the bosonized Hamiltonian, the equivalence of the
``staggered" and ``dimerized" XXZ model is also apparent:
For the dimerized model, the relevant operator is proportional
 to $u\sin{2\varphi}$, while for 
a staggered field, it is $\propto h\cos{2\varphi}$. The phase shift of $\pi /2$
corresponds to the coupling to the bond (in the hopping term) versus coupling to the site.
As a consequence, the ground state for the dimerized model is  a
spin singlet, with $\varphi_{\rm min}=\pi/2$ and $\sigma^z_{n}=0$, while in a staggered magnetic field, an 
antiferromagnetic ground state, with 
$\varphi_{\rm min}=0$ and $\sigma^z_{n}= (-1)^n$, is obtained.
The critical behavior, however, is  
the same for
both models.
 
We numerically verify these predictions and also  determine quantitatively the phase 
boundary between the localized and the delocalized phase. 
In addition to the 
critical point at  $u=0$, $g=2$, physical arguments suggest that the 
delocalized phase should not exceed the point 
$u=1$, $V=-2$:
the transition 
at $V=-2$, being first order, is stable  
against perturbations \cite{Kohmoto81}, and for $\lambda_t u=1$ every second bond is cut and 
therefore the system clearly is localized. Compared with the
random disorder case \cite{SCH}, one is tempted to expect -- 
incorrectly -- that the
delocalized phase is smaller. For the random case, this is due to a stronger 
renormalization of $g$, namely
${{\rm d} g}/{{\rm d} l}=-0.5g^2W$ \cite{Giamarchi88};
compare this with equation \eqref{KTb}.

We will use the phase sensitivity, i.e. the reaction of the system to a change in the
boundary condition, to determine this transition numerically for systems with finite size. 
The ground state energy $E(\phi)$ depends on the boundary condition, expressed via the phase $\phi$.
In particular, we determine below
the energy difference between periodic and anti-periodic boundary conditions,
$\Delta E= (-)^N(E(0)-E(\pi))$. Recall that, 
for a clean system, the ground state energy and the finite size corrections 
can be obtained from the Bethe Ansatz \cite{Eckle87}. 
At half filling (and for odd particle number), the result, in the Luttinger regime,
is given by \cite{Shastry90}
\begin{equation}\label{Bethe}
E_M(\phi)-M\varepsilon_{\infty}=-\frac{\pi v}{6 M}\left(1-3g\frac{\phi^2}{\pi^2}\right),
\end{equation}
where $\varepsilon_{\infty}$ is the 
energy density in the thermodynamic limit. Thus $M\Delta E=\pi v g/2$, independent of $M$, for the 
metallic state. In an insulator, on the other hand, the system cannot react to a twist in the boundary condition,
i.e.\  $M\Delta E$ is expected to decrease with system size.

For non-interacting fermions,  the finite size corrections  for the 
Su-Schrieffer-Heger (SSH)  model \cite{SSH}, $\varepsilon_{\infty}(u) \propto u^2\ln{u}$,  can be determined in an elementary way, with the help of the
Euler-McLaurin
formula. However, it is necessary to handle the van Hove singularity at the band edge
carefully for periodic boundary conditions, while for
anti-periodic boundary conditions,
this term is absent.
With $h=2\pi/M$, we obtain the following result:
\begin{eqnarray}\label{SSH} 
E_M(u)-M\varepsilon_{\infty}(u)&=&\left(\frac{u^2}{2h}\ln{u}-
\frac{u^2}{2h}\ln{(\sqrt{u^2+h^2}-h)} \right. \notag
\\   &-&\left.\frac{1}{2}u +\frac{1}{12}\frac{h^2(1-u^2)}{\sqrt{h^2+u^2}}
-\frac{1}{12}h\right).\end{eqnarray} 
Thus,   for small systems, $h\gg u$, we obtain a correction linear in $u$ which 
does not depend on the system size, and a correction to the $u^2\ln{u}$-term: 
$$
\approx -\frac{1}{2}u-\frac{u^2}{2h}\ln{\frac{u}{2h}},
$$
which we plot in figure \ref{fig1}.
For large systems, $h \ll u$, the  corrections are of order $1/M$:
$$
\approx -\frac{1}{12}h+\frac{1}{12}\frac{h^2}{u}.
$$
Thus, for small  enough systems, or small enough $u$, 
the phase sensitivity is $M\Delta E(u)\propto -uM$; this behavior 
results from the zone  boundary contribution. 

For $V\neq 0$, we combine the treatment of Loss \cite{Loss92} of a 
Luttinger liquid on a closed ring with the
scaling equation \eqref{KTa}, and a 
first order perturbative calculation, with the following result: 
\begin{equation}
\label{stiffu} 
M\Delta E(\lambda u)=M\Delta E(\lambda u=0)-\lambda u \frac{2\pi}{v}\left(\frac{M}{M_0}\right)^{2-g} .
\end{equation}
This is  similar  to  the case of a single impurity \cite{SCH}, but valid in a  restricted regime only.
This estimation 
cannot be extended to large $M$ or $u$, because, in a finite system with a
gap $\Delta (u)$, the ``free motion" is only seen if $L/\xi\propto M \Delta (u)\propto Mu$ 
is small. For larger systems, the phase sensitivity
depends exponentially on the correlation length, as discussed in \cite{baxter}.
In addition,  we have neglected the renormalization of the 
interaction parameter $g$, see equation \eqref{KTb}, which is reasonable only for $V<0$,
as can be seen easily by integrating the RG equations \eqref{KTa} and \eqref{KTb}.   
For example, only  the samples with $u =0.01$,  for system sizes $M=24$ and $M=50$, can be fitted 
with equation \eqref{stiffu}. Already for $u=0.05$ and $M=24$ deviations from the first order treatment are obvious; compare figure \ref{fig2}.
 
Our aim is to determine quantitatively the phase boundary for the dimerized  
XXZ model
($\lambda_V=\lambda_t$), and to compare the results with  those in the  
case of
$\lambda_V=0$. In the latter case, one may compare with results obtained from
the massive Thirring model, or the RG equations.  
In our density matrix renormalization group calculations \cite{White92}, we kept 1000 states and performed 9 finite lattice sweeps.
First, consider $\lambda_V=0$.
In order to  determine the phase transition near $u=0$,
we plot in  figure \ref{fig3} the phase sensitivity for  $M=50$ and 100, and 
small $u$, and confirm that the transition occurs near $V\approx -1.4$.
(The transition point is defined as the point where the phase sensitivity becomes
independent of the system size.)
Proceeding to higher values of $u$, we can follow the phase boundary,
 as shown in figure \ref{fig4}. Already at $u=0.4$, we find a completely localized phase, and we verify the
presumption that the delocalized phase extends to the line $V=-2$.
In figures \ref{fig5} and \ref{fig6}, we show the difference between 
the two cases, $\lambda_V=\lambda_t$ versus
$\lambda_V=0$.
There is no drastic change in the qualitative behavior for small $u$,
and the transition point  remains, as expected from \eqref{boson} and the 
renormalization group equation \eqref{KTa}, at $V\approx -\sqrt{2}$. We also
find, in the case 
$\lambda_V =\lambda_t$, that 
the phase sensitivity decreases more slowly for an attractive interaction, 
and faster for a repulsive interaction.
Increasing the dimerization further, we see that the delocalized phase is
considerably larger for $\lambda_V=\lambda_t$, and indeed extends
to the point $u=1$. In comparison with the disordered model \cite{SCH},
the transition occurs at a stronger interaction, and the delocalized 
region extends to the line $V=-2$.
\section{Interaction and dimerization}
In the last section, we determined the $u$-$V$ phase diagram, 
under the assumption
that the lattice distortion is a fixed parameter. In a more realistic model,
the harmonic lattice energy will compensate
the energy gain of the fermions or the spins, respectively,
resulting in a finite equilibrium value of $u$.  

We consider  the
ground state energy for periodic boundary conditions, studying the lattice
effect for $-2<V<2$ first.
The numerical data (see figures \ref{fig7}
and \ref{fig8}), obtained for 100- or 200-site systems, while keeping  
about 300 to 400 states
(for about 100 sites the finite size corrections are very small and 
the term linear in $u$ is absent), 
show that we can distinguish three regimes. For $V$ between $-2$ and 0,
the gain in the fermionic energy grows parabolically,
$\Delta E(\lambda u)\propto -(\lambda u)^2$. For $0<V<2$, the energy gain
is stronger, $\propto u^\alpha$, with an exponent which reaches its
minimum of 4/3 at $V=2$, where correlations change the behavior.

Explicitly, from an analysis of the massive Thirring model, one finds
\cite{Fuku81}  
the energy gain, $\Delta E(\lambda u)=E(\lambda u)-E(u=0)$, to be given by 
\begin{eqnarray}\label{Eferm}
 -\sqrt{2}<V \! <0 &: \;\; & \Delta E(\lambda u)=c_1u^\alpha-c_2u^2,\nonumber\\
&& \alpha=2/(2-g)>2; \nonumber  \\
0<V\leq 2 &: \;\; & \Delta E(\lambda u)= -c_1u^\alpha+c_2u^2, \nonumber \\ &&  
4/3<
\alpha <2,
\end{eqnarray}
with $c_1 > c_2 >0$; compare figure 9.
The non-quadratic term can also be obtained from 
a self-consistent treatment of the perturbation. In this so-called
self-consistent harmonic approximation (SCHA) \cite{Dashen75,Fukuyama85}
an optimal
quadratic approximation for the periodic potential 
around one of its minima is chosen, which 
corresponds to the following replacement: 
\begin{eqnarray}
U\cos{n\varphi} \to U\big[1-\frac{n^2}{2}(\varphi^2-\langle\varphi^2\rangle)\big]
e^{-n^2\langle\varphi^2\rangle/2} ;
\end{eqnarray} 
here, explicitly,
\begin{eqnarray}
\langle\varphi^2\rangle=\frac{g}{2}
\ln{\frac{2\pi}{l_0q_0}}, \quad q^2_0=\frac{ngU}{v} e^{-n^2\langle\varphi^2\rangle/2} = \frac{\Delta ^2}{v^2}.
\end{eqnarray}
The quadratic term in equation (\ref{Eferm}) is related to 
the tunneling between the minima of the 
periodic potential \cite{Gomez96}.
The  logarithmic dependence on $u$ for $V=0$, which was found in \cite {SSH},
is contained in the above formulas:
\begin{eqnarray}\label{V0}
V=0 &:& \Delta E_0 =-\frac{Mt}{\pi}(\lambda u)^2(\ln{\frac{4}{\lambda u}}-0.5) \nonumber \\
V\approx 0 &:& \Delta E =\Delta E_0 -\frac{MV}{\pi}(\lambda u)^2
\ln^2{\frac{4}{\lambda u}}. 
\end{eqnarray}
Adding the harmonic lattice energy, $MK_0 u^2 /2$, we realize that a finite
equilibrium dimerization cannot be be stabilized for an attractive 
interaction. For small $|V|$, 
the second logarithmic term in (\ref{V0}) always leads to a positive
slope  of $\Delta E(\lambda u)$ at $u=0$; and for large (negative) $|V|$, 
we find, as shown in the previous section, a delocalized phase for small $u$. 
   
For the non-interacting case ($V=0$), the stable dimerization $\lambda u_0$
and the energy gain are given by well-known expressions \cite{Fukuyama85}
for CDW-systems. With $\gamma=\lambda ^2t/K_0$, the results are
\begin{eqnarray}\label{u00}
\lambda u_0 &\propto & {\rm e}^{-1/\gamma} , \nonumber \\
\Delta (\lambda u_0) &\propto & {\rm e}^{-1/\gamma} , \\
E_{\rm total}(\lambda u_0) &\propto &  {\rm e}^{-2/\gamma} . \nonumber
\end{eqnarray}
For a repulsive interaction, we find the following algebraic dependences: 
\begin{eqnarray}\label{u0V}
\lambda u_0 &\propto & \gamma ^{\frac{2-g}{2-2g}} , \nonumber \\
\Delta (\lambda u_0) &\propto & \gamma^{\frac{1}{2(1-g)}} , \\
E_{\rm total}(\lambda u_0) &\propto & \gamma^{\frac{1}{1-g}} , \nonumber
\end{eqnarray}
and, for example,
$$
\begin{array}{ccc}
V=1: \;\; &\lambda u_0\propto \gamma^{5/2}, \quad &
E_{\rm total}(\lambda u_0)\propto \gamma^4 ;\\
V=2: \;\; &\lambda u_0\propto \gamma^{3/2},   &
E_{\rm total}(\lambda u_0)\propto \gamma ^2.
\end{array}$$
Another interesting quantity is the curvature of the total energy 
at the minimum, which is related to the zone boundary phonon frequency,
$\omega_{2k_F}$. For a decoupled system ($\lambda =0$), the curvature
is given by $K_0$, the bare value. The fermion-lattice coupling changes this quantity as follows:
\begin{eqnarray}\label{K0}
V=0 & : \;\; & E_{total}''(u_0)/M=\gamma K_0 \nonumber \\
V>0 & : \;\; & E_{total}''(u_0)/M=K_0(2-\alpha ) .
\end{eqnarray}
We thus conclude that for interacting fermions, the softening of the phonon at the zone boundary does not depend on the 
coupling strength $\gamma$, but on the exponent of the
energy gain of the fermions only. 

Next we  consider the regime $V>2$, where a correlation gap develops, for the ``clean" case given by \cite{Shankar}
\begin{eqnarray} \label{gap}
V\gtrsim 2 &: \;\; & \Delta(V)\propto e^{-\pi^2/\sqrt{V-2}}, \nonumber \\
                 V \to \infty &: \;\; & \Delta(V)\propto V .
\end{eqnarray}
A gap also develops in a frustrated
model, i.e.\ when next-nearest-neigh\-bor coupling is included, for
$V=2$ and provided that $V_{\rm frust}>V_{\rm frust}^c$. This is a
consequence of a relevant 4$k_F$-back\-scattering process, and
believed to be relevant for CuGeO$_3$ \cite{Castilla,Riera}.
As shown in \cite{Haldane82}, a
$\cos{4\varphi}$-term leads to a Neel (Ising-type) or a dimer (frustration) ground
state, respectively, depending on its sign.
Thus we expect that a next-nearest-neighbor
interaction will increase the
dimerization, while an Ising-type interaction reduces it.
Possible distinctive features between the frustration induced
and the externally,  by $\lambda u$, driven dimerized state  are discussed
in \cite{Chitra97}.

Considering  the numerical data for a strong repulsive interaction, $V>2$,
see figure \ref{fig8}, 
the energy gain is nearly constant at first, when increasing $V$; 
only the prefactor is slightly reduced. 
Arguments supporting this observation are given below.
Increasing the interaction, 
the energy gain is drastically reduced, the exponent approaches again 2,
and therefore no minimum is found in the total energy.
To understand this behavior, we  
consider in more detail the
extrema of the combined $\sin$-$\cos$-potential,
$U(\varphi)=-\tilde V\cos{4\varphi}
-\tilde u\sin{2\varphi}$, 
where  $\tilde u=-2\pi v g \lambda u/l_0$, and $\tilde V=\pi v g V/l^2_0$. 
For a large interaction, the minima are at 
$\sin 2\varphi_0 = \tilde u /(4{\tilde V})$,
and the expansion reads 
\begin{equation}\label{Vgross}
U(\varphi)\approx -\tilde V-\frac{{\tilde u}^2}{8\tilde V}
+\left(8\tilde V-\frac{{\tilde u}^2}{2\tilde V}\right)(\varphi-\varphi_0)^2
\;\; . 
\end{equation}
So for a strong interaction, it is possible to
consider the $u^2$-terms in the  potential as a small perturbation to the
exact results \cite{Shankar}.
Note that, for a strong interaction (i.e. the renormalized $g$ 
is $\approx 0$), fluctuations are not
important since
$\langle\varphi^2\rangle\approx 0$. These arguments support the result, obtained
 numerically,
that the energy gain is quadratic in $u$.
For large $u$, on the other hand, the minima are located at
$\cos 2\varphi_0 = 0$,
and the expansion for small interaction reads
\begin{equation}\label{ugross}
U(\varphi)\approx (\tilde V-\tilde u)+2(\tilde u-4\tilde V)(\varphi-\varphi_0)^2
 \;\; .
\end{equation}
For an interaction $V\approx 2$, where the effective influence of the 
$\tilde V\cos{4\varphi}$-term is 
expected to be small, we repeat the perturbation analysis, i.e.,  
we evaluate first the $\tilde u\sin{2\varphi}$-term
within the SCHA; thus $\Delta(u) \propto u^{2/3}$.
Adding then the interaction term to the gapped system, and taking into account the 
fluctuations, 
the dimerization gap is reduced:
\begin{equation}\label{24}
\Delta^2(u)\to \frac{\Delta^2(u)}{ (1+\overline{V})}, \quad \overline{V}=\frac{V}{\pi v}. 
\end{equation}
For a frustrating interaction, there is no change in the minima, and 
we find, for a small frustration, a similar expression to \eqref{24}, and 
for a stronger frustration, a linear increase in $u$. 

To check the validity of the perturbation analysis at $V\to2$, we compare with results from the SCHA and from field theory.
In the evaluation of the SCHA scheme, 
 which considers both perturbations on an equal footing following \eqref{ugross},
we find the following gap equation, which  can be solved for $g=1/2$:
\begin{equation}\label{V2}
\Delta^2 = 
 \tilde u \left(\frac{l_0\Delta}{2\pi v }\right)^g -4\tilde V \left(\frac{l_0 
\Delta}{2\pi v}\right)^{4g}; 
\end{equation}
the result is:
\begin{equation}\label{V2b} 
\Delta(u, \overline{V} )=\frac{\Delta(u)}{(1+ \overline{V})^{2/3}}.
\end{equation} 
Clearly, the exponent of the denominator does not coincide 
with the one in \eqref{24}.

However, a more careful treatment, particularly with respect to  the marginality of $\cos{4\varphi}$ at $V=2$,  
introduces a logarithmic correction to the dimerization gap \cite{Schulz96}, given by
\begin{equation}\label{log}
\Delta(u,\overline{V})\propto \frac{\Delta (u)}{[1+ \overline{V}\ln{( v/\Delta (u))}]^{1/2}}. 
\end{equation} 
This logarithmic term is  a consequence of the 
scaling of the  marginal operator
$V(L)=V(1+\pi V\ln{L})^{-1}$, see \cite{Affleck89}, 
and therefore is not found within both of the above considered approximations.  
The remaining difference between \eqref{V2b} and \eqref{log}, namely the different
 exponents, can be explained by higher order contributions 
\cite{Affleck89}.
Thus the perturbation approach, which led to \eqref{24}, is rather close to the correct result, \eqref{log}, and even the simple SCHA is not too far off, 
as it shows that the interaction essentially changes the prefactor, i.e. the gap is $\sim \Delta(u)$. 

The observation that a repulsive interaction in the one-dimensional 
Hubbard model  
can enhance the dimerization \cite{Waas90} can be explained now in the 
following way: 
For small $U$, where the charge gap is small, we find
an energy gain 
$E_{\rm charge}\propto E_{\rm spin}\propto tu^{4/(2-g_{c})}$,  with the corresponding interaction parameters
$g_s=1$ for the spin and $g_c$ for the charge.
At half filling, an intermediate $U$ leads to a large enough gap to push down the dimerization
in the charge channel, but in the spin channel, no gap appears, 
and the system gains
energy from the coupling to the lattice. The charge gap reduces, nevertheless,
the prefactor from $t$ to $t^2/U$.
\section{Summary}
In summary, we have shown that a periodic distortion of the lattice is irrelevant for $V<-\sqrt{2}$, 
resulting in a delocalized phase. 
The transition point, $V_c (u=0)=-\sqrt{2}$, is not modified by the way of 
coupling to the lattice.   
However, the extension of the delocalized phase depends strongly 
on the coupling to the interaction, i.e.\ the parameter 
$\lambda_V$. For the spin model
($\lambda_V=\lambda_t$),
the delocalized region is considerably larger than for the case $\lambda_V =0$.
In comparison with a random disorder model \cite{SCH}, the most important difference 
is the fact that the delocalized region extends down to the line $V=-2$. 
The phase diagram is summarized in figure 
\ref{fig10}. Due to the difficulties in determining precisely 
the point where the curves for different system sizes coincide, 
we can plot only a rough phase diagramr; nevertheless the 
transition seems to be steeper for $\lambda_V=\lambda_t$,
compared to the case $\lambda_V=0$.

Including the harmonic lattice energy, 
we find that the dimerized state is stable, i.e.\ the  total energy develops a minimum at a finite $u_0$,  for a repulsive interaction only. 
With increasing interaction, the dimerization increases even further,
provided no interaction induced, competing ordering occurs.
A schematic plot is given in figure \ref{fig11}. Furthermore, already at
$V=4$, no equilibrium dimerization develops,
 as can be seen from the preliminary data shown 
in figure \ref{fig12}, where we plot the gap versus inverse system size. 
(The lines are obtained from a polynomial
fit in $1/M$ for the case $u=0$, and an exponential fit as mentioned
in \cite{Bouzerar98}, respectively.)
Obviously, no additional gap from the 
dimerization is seen for $V=4$. In contrast, for   smaller $V$, especially $V=2$, 
the dimerization contribution is significant. 
\begin{acknowledgement}
We thank Peter Schmitteckert for providing us with the DMRG algorithm,
and Peter Schwab for useful discussions.
 This work was supported by the Deutsche Forschungsgemeinschaft
(For\-scher\-grup\-pe HO 955/2-1). The calculations were mostly done on the IBM
SP2 at the Leibniz-Rechenzentrum in Munich.
\end{acknowledgement}
%

\setlength{\unitlength}{0.1bp}
\special{!
/gnudict 40 dict def
gnudict begin
/Color false def
/Solid false def
/gnulinewidth 5.000 def
/vshift -33 def
/dl {10 mul} def
/hpt 31.5 def
/vpt 31.5 def
/M {moveto} bind def
/L {lineto} bind def
/R {rmoveto} bind def
/V {rlineto} bind def
/vpt2 vpt 2 mul def
/hpt2 hpt 2 mul def
/Lshow { currentpoint stroke M
  0 vshift R show } def
/Rshow { currentpoint stroke M
  dup stringwidth pop neg vshift R show } def
/Cshow { currentpoint stroke M
  dup stringwidth pop -2 div vshift R show } def
/DL { Color {setrgbcolor Solid {pop []} if 0 setdash }
 {pop pop pop Solid {pop []} if 0 setdash} ifelse } def
/BL { stroke gnulinewidth 2 mul setlinewidth } def
/AL { stroke gnulinewidth 2 div setlinewidth } def
/PL { stroke gnulinewidth setlinewidth } def
/LTb { BL [] 0 0 0 DL } def
/LTa { AL [1 dl 2 dl] 0 setdash 0 0 0 setrgbcolor } def
/LT0 { PL [] 0 1 0 DL } def
/LT1 { PL [4 dl 2 dl] 0 0 1 DL } def
/LT2 { PL [2 dl 3 dl] 1 0 0 DL } def
/LT3 { PL [1 dl 1.5 dl] 1 0 1 DL } def
/LT4 { PL [5 dl 2 dl 1 dl 2 dl] 0 1 1 DL } def
/LT5 { PL [4 dl 3 dl 1 dl 3 dl] 1 1 0 DL } def
/LT6 { PL [2 dl 2 dl 2 dl 4 dl] 0 0 0 DL } def
/LT7 { PL [2 dl 2 dl 2 dl 2 dl 2 dl 4 dl] 1 0.3 0 DL } def
/LT8 { PL [2 dl 2 dl 2 dl 2 dl 2 dl 2 dl 2 dl 4 dl] 0.5 0.5 0.5 DL } def
/P { stroke [] 0 setdash
  currentlinewidth 2 div sub M
  0 currentlinewidth V stroke } def
/D { stroke [] 0 setdash 2 copy vpt add M
  hpt neg vpt neg V hpt vpt neg V
  hpt vpt V hpt neg vpt V closepath stroke
  P } def
/A { stroke [] 0 setdash vpt sub M 0 vpt2 V
  currentpoint stroke M
  hpt neg vpt neg R hpt2 0 V stroke
  } def
/B { stroke [] 0 setdash 2 copy exch hpt sub exch vpt add M
  0 vpt2 neg V hpt2 0 V 0 vpt2 V
  hpt2 neg 0 V closepath stroke
  P } def
/C { stroke [] 0 setdash exch hpt sub exch vpt add M
  hpt2 vpt2 neg V currentpoint stroke M
  hpt2 neg 0 R hpt2 vpt2 V stroke } def
/T { stroke [] 0 setdash 2 copy vpt 1.12 mul add M
  hpt neg vpt -1.62 mul V
  hpt 2 mul 0 V
  hpt neg vpt 1.62 mul V closepath stroke
  P  } def
/S { 2 copy A C} def
end
}
\begin{picture}(2519,1511)(0,0)
\special{"
gnudict begin
gsave
50 50 translate
0.100 0.100 scale
0 setgray
/Helvetica findfont 100 scalefont setfont
newpath
-500.000000 -500.000000 translate
LTa
LTb
600 251 M
63 0 V
1673 0 R
-63 0 V
600 385 M
63 0 V
1673 0 R
-63 0 V
600 520 M
63 0 V
1673 0 R
-63 0 V
600 654 M
63 0 V
1673 0 R
-63 0 V
600 788 M
63 0 V
1673 0 R
-63 0 V
600 923 M
63 0 V
1673 0 R
-63 0 V
600 1057 M
63 0 V
1673 0 R
-63 0 V
600 1191 M
63 0 V
1673 0 R
-63 0 V
600 1326 M
63 0 V
1673 0 R
-63 0 V
600 1460 M
63 0 V
1673 0 R
-63 0 V
874 251 M
0 63 V
0 1146 R
0 -63 V
1240 251 M
0 63 V
0 1146 R
0 -63 V
1605 251 M
0 63 V
0 1146 R
0 -63 V
1971 251 M
0 63 V
0 1146 R
0 -63 V
2336 251 M
0 63 V
0 1146 R
0 -63 V
600 251 M
1736 0 V
0 1209 V
-1736 0 V
600 251 L
LT0
1396 520 D
600 1425 D
691 1388 D
783 1349 D
874 1308 D
965 1264 D
1057 1218 D
1148 1169 D
1240 1119 D
1331 1066 D
1422 1011 D
1514 954 D
1605 894 D
1696 833 D
1788 769 D
1879 703 D
1971 635 D
2062 564 D
2153 492 D
2245 417 D
2336 340 D
LT1
1396 420 A
600 1458 A
691 1453 A
783 1443 A
874 1430 A
965 1413 A
1057 1393 A
1148 1369 A
1240 1341 A
1331 1310 A
1422 1275 A
1514 1237 A
1605 1196 A
1696 1151 A
1788 1102 A
1879 1051 A
1971 997 A
2062 939 A
2153 878 A
2245 815 A
2336 748 A
LT2
600 1425 M
18 -7 V
17 -7 V
18 -7 V
17 -7 V
18 -7 V
17 -7 V
18 -8 V
17 -7 V
18 -8 V
17 -7 V
18 -8 V
17 -8 V
18 -8 V
17 -8 V
18 -8 V
18 -8 V
17 -8 V
18 -8 V
17 -9 V
18 -8 V
17 -9 V
18 -8 V
17 -9 V
18 -9 V
17 -8 V
18 -9 V
17 -9 V
18 -9 V
18 -10 V
17 -9 V
18 -9 V
17 -10 V
18 -9 V
17 -10 V
18 -9 V
17 -10 V
18 -10 V
17 -10 V
18 -10 V
17 -10 V
18 -10 V
17 -10 V
18 -10 V
18 -11 V
17 -10 V
18 -11 V
17 -10 V
18 -11 V
17 -11 V
18 -11 V
17 -10 V
18 -11 V
17 -12 V
18 -11 V
17 -11 V
18 -11 V
18 -12 V
17 -11 V
18 -12 V
17 -11 V
18 -12 V
17 -12 V
18 -12 V
17 -12 V
18 -12 V
17 -12 V
18 -12 V
17 -13 V
18 -12 V
17 -12 V
18 -13 V
18 -13 V
17 -12 V
18 -13 V
17 -13 V
18 -13 V
17 -13 V
18 -13 V
17 -13 V
18 -13 V
17 -14 V
18 -13 V
17 -14 V
18 -13 V
18 -14 V
17 -14 V
18 -14 V
17 -13 V
18 -14 V
17 -14 V
18 -15 V
17 -14 V
18 -14 V
17 -15 V
18 -14 V
17 -15 V
18 -14 V
17 -15 V
18 -15 V
LT3
600 1456 M
18 -1 V
17 -2 V
18 -2 V
17 -2 V
18 -2 V
17 -2 V
18 -3 V
17 -2 V
18 -3 V
17 -3 V
18 -4 V
17 -3 V
18 -4 V
17 -3 V
18 -4 V
18 -4 V
17 -5 V
18 -4 V
17 -5 V
18 -4 V
17 -5 V
18 -5 V
17 -5 V
18 -6 V
17 -5 V
18 -6 V
17 -5 V
18 -6 V
18 -6 V
17 -6 V
18 -6 V
17 -7 V
18 -6 V
17 -7 V
18 -7 V
17 -7 V
18 -7 V
17 -7 V
18 -7 V
17 -8 V
18 -7 V
17 -8 V
18 -8 V
18 -8 V
17 -8 V
18 -8 V
17 -8 V
18 -9 V
17 -8 V
18 -9 V
17 -9 V
18 -9 V
17 -9 V
18 -9 V
17 -9 V
18 -10 V
18 -9 V
17 -10 V
18 -9 V
17 -10 V
18 -10 V
17 -10 V
18 -10 V
17 -11 V
18 -10 V
17 -11 V
18 -10 V
17 -11 V
18 -11 V
17 -11 V
18 -11 V
18 -11 V
17 -11 V
18 -11 V
17 -12 V
18 -11 V
17 -12 V
18 -12 V
17 -12 V
18 -12 V
17 -12 V
18 -12 V
17 -12 V
18 -13 V
18 -12 V
17 -13 V
18 -12 V
17 -13 V
18 -13 V
17 -13 V
18 -13 V
17 -13 V
18 -13 V
17 -14 V
18 -13 V
17 -14 V
18 -13 V
17 -14 V
18 -14 V
stroke
grestore
end
showpage
}
\put(1276,420){\makebox(0,0)[r]{anti-periodic}}
\put(1276,520){\makebox(0,0)[r]{periodic }}
\put(1468,51){\makebox(0,0){$\lambda u$}}
\put(100,855){%
\special{ps: gsave currentpoint currentpoint translate
270 rotate neg exch neg exch translate}%
\makebox(0,0)[b]{\shortstack{$\Delta E( \lambda u)$}}%
\special{ps: currentpoint grestore moveto}%
}
\put(2336,151){\makebox(0,0){0.05}}
\put(1971,151){\makebox(0,0){0.04}}
\put(1605,151){\makebox(0,0){0.03}}
\put(1240,151){\makebox(0,0){0.02}}
\put(874,151){\makebox(0,0){0.01}}
\put(540,1460){\makebox(0,0)[r]{0}}
\put(540,1326){\makebox(0,0)[r]{-0.02}}
\put(540,1191){\makebox(0,0)[r]{-0.04}}
\put(540,1057){\makebox(0,0)[r]{-0.06}}
\put(540,923){\makebox(0,0)[r]{-0.08}}
\put(540,788){\makebox(0,0)[r]{-0.1}}
\put(540,654){\makebox(0,0)[r]{-0.12}}
\put(540,520){\makebox(0,0)[r]{-0.14}}
\put(540,385){\makebox(0,0)[r]{-0.16}}
\put(540,251){\makebox(0,0)[r]{-0.18}}
\end{picture}
 FIG. \ref{fig1}: \refstepcounter{figure} \label{fig1}
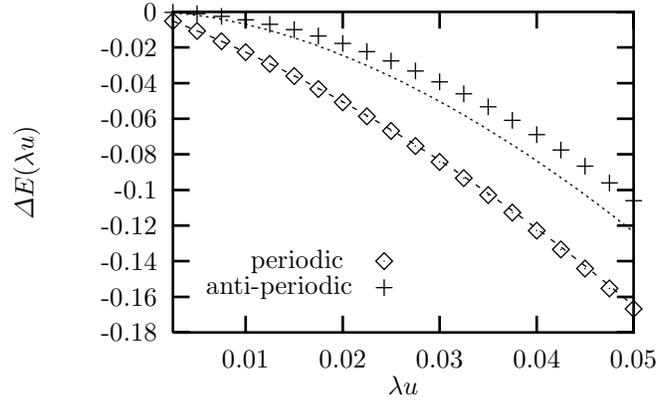
Finite size corrections to the ground state energy for $V=0$.
The points (${\diamond\!\negmedspace\cdot}$, $+$) are numerical data for a 40-site system, with periodic or anti-periodic boundary conditions.
The dotted line shows the ground state energy in the 
thermodynamic limit, $\propto Mu^2\ln{u}$,
the dashed line is the energy with the corrections according 
to equation \eqref{SSH}. 
\\  \\ 
\setlength{\unitlength}{0.1bp}
\special{!
/gnudict 40 dict def
gnudict begin
/Color false def
/Solid false def
/gnulinewidth 5.000 def
/vshift -33 def
/dl {10 mul} def
/hpt 31.5 def
/vpt 31.5 def
/M {moveto} bind def
/L {lineto} bind def
/R {rmoveto} bind def
/V {rlineto} bind def
/vpt2 vpt 2 mul def
/hpt2 hpt 2 mul def
/Lshow { currentpoint stroke M
  0 vshift R show } def
/Rshow { currentpoint stroke M
  dup stringwidth pop neg vshift R show } def
/Cshow { currentpoint stroke M
  dup stringwidth pop -2 div vshift R show } def
/DL { Color {setrgbcolor Solid {pop []} if 0 setdash }
 {pop pop pop Solid {pop []} if 0 setdash} ifelse } def
/BL { stroke gnulinewidth 2 mul setlinewidth } def
/AL { stroke gnulinewidth 2 div setlinewidth } def
/PL { stroke gnulinewidth setlinewidth } def
/LTb { BL [] 0 0 0 DL } def
/LTa { AL [1 dl 2 dl] 0 setdash 0 0 0 setrgbcolor } def
/LT0 { PL [] 0 1 0 DL } def
/LT1 { PL [4 dl 2 dl] 0 0 1 DL } def
/LT2 { PL [2 dl 3 dl] 1 0 0 DL } def
/LT3 { PL [1 dl 1.5 dl] 1 0 1 DL } def
/LT4 { PL [5 dl 2 dl 1 dl 2 dl] 0 1 1 DL } def
/LT5 { PL [4 dl 3 dl 1 dl 3 dl] 1 1 0 DL } def
/LT6 { PL [2 dl 2 dl 2 dl 4 dl] 0 0 0 DL } def
/LT7 { PL [2 dl 2 dl 2 dl 2 dl 2 dl 4 dl] 1 0.3 0 DL } def
/LT8 { PL [2 dl 2 dl 2 dl 2 dl 2 dl 2 dl 2 dl 4 dl] 0.5 0.5 0.5 DL } def
/P { stroke [] 0 setdash
  currentlinewidth 2 div sub M
  0 currentlinewidth V stroke } def
/D { stroke [] 0 setdash 2 copy vpt add M
  hpt neg vpt neg V hpt vpt neg V
  hpt vpt V hpt neg vpt V closepath stroke
  P } def
/A { stroke [] 0 setdash vpt sub M 0 vpt2 V
  currentpoint stroke M
  hpt neg vpt neg R hpt2 0 V stroke
  } def
/B { stroke [] 0 setdash 2 copy exch hpt sub exch vpt add M
  0 vpt2 neg V hpt2 0 V 0 vpt2 V
  hpt2 neg 0 V closepath stroke
  P } def
/C { stroke [] 0 setdash exch hpt sub exch vpt add M
  hpt2 vpt2 neg V currentpoint stroke M
  hpt2 neg 0 R hpt2 vpt2 V stroke } def
/T { stroke [] 0 setdash 2 copy vpt 1.12 mul add M
  hpt neg vpt -1.62 mul V
  hpt 2 mul 0 V
  hpt neg vpt 1.62 mul V closepath stroke
  P  } def
/S { 2 copy A C} def
end
}
\begin{picture}(2519,1511)(0,0)
\special{"
gnudict begin
gsave
50 50 translate
0.100 0.100 scale
0 setgray
/Helvetica findfont 100 scalefont setfont
newpath
-500.000000 -500.000000 translate
LTa
LTb
600 251 M
63 0 V
1673 0 R
-63 0 V
600 372 M
63 0 V
1673 0 R
-63 0 V
600 493 M
63 0 V
1673 0 R
-63 0 V
600 614 M
63 0 V
1673 0 R
-63 0 V
600 735 M
63 0 V
1673 0 R
-63 0 V
600 856 M
63 0 V
1673 0 R
-63 0 V
600 976 M
63 0 V
1673 0 R
-63 0 V
600 1097 M
63 0 V
1673 0 R
-63 0 V
600 1218 M
63 0 V
1673 0 R
-63 0 V
600 1339 M
63 0 V
1673 0 R
-63 0 V
600 1460 M
63 0 V
1673 0 R
-63 0 V
600 251 M
0 63 V
0 1146 R
0 -63 V
1034 251 M
0 63 V
0 1146 R
0 -63 V
1468 251 M
0 63 V
0 1146 R
0 -63 V
1902 251 M
0 63 V
0 1146 R
0 -63 V
2336 251 M
0 63 V
0 1146 R
0 -63 V
600 251 M
1736 0 V
0 1209 V
-1736 0 V
600 251 L
LT0
618 1083 M
17 26 V
18 19 V
17 16 V
18 14 V
17 12 V
18 11 V
17 10 V
18 10 V
17 9 V
18 8 V
17 8 V
18 7 V
17 7 V
18 7 V
18 6 V
17 6 V
18 6 V
17 5 V
18 6 V
17 5 V
18 5 V
17 5 V
18 4 V
17 5 V
18 4 V
17 4 V
18 4 V
18 4 V
17 4 V
18 4 V
17 3 V
18 4 V
17 3 V
18 3 V
17 4 V
18 3 V
17 3 V
18 3 V
17 2 V
18 3 V
17 3 V
18 2 V
18 3 V
17 3 V
18 2 V
17 2 V
18 3 V
17 2 V
18 2 V
17 2 V
18 2 V
17 2 V
18 2 V
17 2 V
18 2 V
18 1 V
17 2 V
18 2 V
17 1 V
18 2 V
17 1 V
18 2 V
17 1 V
18 2 V
17 1 V
18 1 V
17 2 V
18 1 V
17 1 V
18 1 V
18 1 V
17 1 V
18 1 V
17 1 V
18 1 V
17 1 V
18 0 V
17 1 V
18 1 V
17 1 V
18 0 V
17 1 V
18 1 V
18 0 V
17 1 V
18 0 V
17 0 V
18 1 V
17 0 V
18 1 V
17 0 V
18 0 V
17 0 V
18 0 V
17 1 V
18 0 V
17 0 V
18 0 V
LT1
2336 1425 D
1902 1414 D
1468 1376 D
1294 1352 D
1208 1337 D
1121 1319 D
1086 1310 D
1034 1299 D
947 1275 D
860 1246 D
774 1209 D
687 1158 D
643 1119 D
LT2
1588 553 A
2336 887 A
1902 1057 A
1468 1202 A
1294 1241 A
1208 1254 A
1121 1261 A
1103 1262 A
1034 1262 A
947 1254 A
860 1237 A
774 1207 A
687 1157 A
LT3
618 1083 M
17 26 V
18 19 V
17 16 V
18 14 V
17 12 V
18 11 V
17 10 V
18 9 V
17 9 V
18 8 V
17 7 V
18 7 V
17 6 V
18 6 V
18 5 V
17 4 V
18 5 V
17 3 V
18 4 V
17 3 V
18 2 V
17 2 V
18 2 V
17 1 V
18 1 V
17 1 V
18 0 V
18 0 V
17 -1 V
18 0 V
17 -1 V
18 -2 V
17 -1 V
18 -2 V
17 -2 V
18 -2 V
17 -3 V
18 -2 V
17 -3 V
18 -3 V
17 -4 V
18 -3 V
18 -4 V
17 -4 V
18 -4 V
17 -4 V
18 -5 V
17 -4 V
18 -5 V
17 -5 V
18 -5 V
17 -5 V
18 -5 V
17 -5 V
18 -6 V
18 -5 V
17 -6 V
18 -6 V
17 -5 V
18 -6 V
17 -6 V
18 -7 V
17 -6 V
18 -6 V
17 -6 V
18 -7 V
17 -6 V
18 -7 V
17 -7 V
18 -6 V
18 -7 V
17 -7 V
18 -7 V
17 -7 V
18 -7 V
17 -7 V
18 -8 V
17 -7 V
18 -7 V
17 -7 V
18 -8 V
17 -7 V
18 -8 V
18 -7 V
17 -8 V
18 -8 V
17 -7 V
18 -8 V
17 -8 V
18 -8 V
17 -8 V
18 -7 V
17 -8 V
18 -8 V
17 -8 V
18 -9 V
17 -8 V
18 -8 V
LT4
1588 453 B
1468 1257 B
1294 1266 B
1121 1265 B
1034 1260 B
774 1205 B
687 1160 B
600 1067 B
LT5
618 1083 M
17 26 V
18 19 V
17 16 V
18 14 V
17 12 V
18 11 V
17 10 V
18 9 V
17 8 V
18 7 V
17 7 V
18 7 V
17 5 V
18 6 V
18 4 V
17 5 V
18 4 V
17 3 V
18 3 V
17 3 V
18 3 V
17 2 V
18 2 V
17 2 V
18 1 V
17 2 V
18 1 V
18 1 V
17 1 V
18 0 V
17 1 V
18 0 V
17 0 V
18 0 V
17 0 V
18 0 V
17 -1 V
18 0 V
17 -1 V
18 0 V
17 -1 V
18 -1 V
18 0 V
17 -1 V
18 -1 V
17 -1 V
18 -2 V
17 -1 V
18 -1 V
17 -1 V
18 -2 V
17 -1 V
18 -2 V
17 -1 V
18 -2 V
18 -1 V
17 -2 V
18 -1 V
17 -2 V
18 -2 V
17 -2 V
18 -1 V
17 -2 V
18 -2 V
17 -2 V
18 -2 V
17 -2 V
18 -2 V
17 -2 V
18 -2 V
18 -2 V
17 -2 V
18 -2 V
17 -2 V
18 -3 V
17 -2 V
18 -2 V
17 -2 V
18 -2 V
17 -3 V
18 -2 V
17 -2 V
18 -3 V
18 -2 V
17 -3 V
18 -2 V
17 -2 V
18 -3 V
17 -2 V
18 -3 V
17 -2 V
18 -3 V
17 -3 V
18 -2 V
17 -3 V
18 -2 V
17 -3 V
18 -3 V
LT6
1588 353 C
1468 802 C
1294 915 C
1121 1025 C
1034 1075 C
774 1161 C
687 1143 C
600 1061 C
LT7
618 1083 M
17 26 V
18 19 V
17 16 V
18 13 V
17 12 V
18 10 V
17 9 V
18 7 V
17 5 V
18 4 V
17 3 V
18 2 V
17 0 V
18 0 V
18 -2 V
17 -3 V
18 -4 V
17 -4 V
18 -6 V
17 -6 V
18 -7 V
17 -7 V
18 -8 V
17 -9 V
18 -9 V
17 -9 V
18 -10 V
18 -10 V
17 -11 V
18 -11 V
17 -11 V
18 -11 V
17 -12 V
18 -12 V
17 -12 V
18 -12 V
17 -12 V
18 -13 V
17 -13 V
18 -12 V
17 -13 V
18 -13 V
18 -13 V
17 -13 V
18 -12 V
17 -13 V
18 -13 V
17 -14 V
18 -13 V
17 -13 V
18 -13 V
17 -13 V
18 -13 V
17 -13 V
18 -13 V
18 -13 V
17 -13 V
18 -13 V
17 -12 V
18 -13 V
17 -13 V
18 -13 V
17 -13 V
18 -12 V
17 -13 V
18 -13 V
17 -12 V
18 -13 V
17 -12 V
18 -13 V
18 -12 V
17 -12 V
18 -13 V
17 -12 V
18 -12 V
17 -12 V
18 -12 V
17 -13 V
18 -12 V
17 -12 V
18 -12 V
17 -11 V
18 -12 V
18 -12 V
17 -12 V
18 -12 V
17 -11 V
18 -12 V
17 -12 V
18 -11 V
17 -12 V
18 -11 V
17 -12 V
18 -11 V
17 -12 V
18 -11 V
17 -12 V
18 -11 V
stroke
grestore
end
showpage
}
\put(1468,353){\makebox(0,0)[r]{$u=0.05$, $M=24$}}
\put(1468,453){\makebox(0,0)[r]{$u=0.01$, $M=24$}}
\put(1468,553){\makebox(0,0)[r]{$u=0.01$, $M=50$}}
\put(1468,51){\makebox(0,0){$V$}}
\put(100,855){%
\special{ps: gsave currentpoint currentpoint translate
270 rotate neg exch neg exch translate}%
\makebox(0,0)[b]{\shortstack{$M\Delta E(u)$}}%
\special{ps: currentpoint grestore moveto}%
}
\put(2336,151){\makebox(0,0){0}}
\put(1902,151){\makebox(0,0){-0.5}}
\put(1468,151){\makebox(0,0){-1}}
\put(1034,151){\makebox(0,0){-1.5}}
\put(600,151){\makebox(0,0){-2}}
\put(540,1460){\makebox(0,0)[r]{3.2}}
\put(540,1339){\makebox(0,0)[r]{3}}
\put(540,1218){\makebox(0,0)[r]{2.8}}
\put(540,1097){\makebox(0,0)[r]{2.6}}
\put(540,976){\makebox(0,0)[r]{2.4}}
\put(540,856){\makebox(0,0)[r]{2.2}}
\put(540,735){\makebox(0,0)[r]{2}}
\put(540,614){\makebox(0,0)[r]{1.8}}
\put(540,493){\makebox(0,0)[r]{1.6}}
\put(540,372){\makebox(0,0)[r]{1.4}}
\put(540,251){\makebox(0,0)[r]{1.2}}
\end{picture}
 FIG. \ref{fig2}: \refstepcounter{figure} \label{fig2}
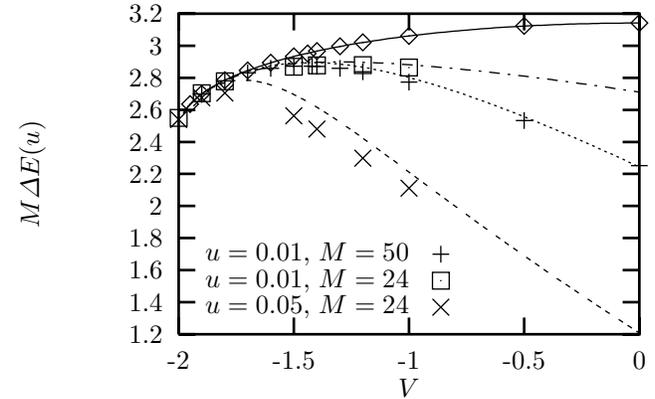
Comparison of  numerical data with equation \eqref{stiffu}; $\lambda_t=1$, $\lambda_V=0$, 
$M_0\approx 2$. 
The  $\diamond$ are numerical data for $u=0$ and $M=50$, 
in apparent agreement with the analytical result, equation \eqref{Bethe} 
(full line). 
\\   \\ 
\setlength{\unitlength}{0.1bp}
\special{!
/gnudict 40 dict def
gnudict begin
/Color false def
/Solid false def
/gnulinewidth 5.000 def
/vshift -33 def
/dl {10 mul} def
/hpt 31.5 def
/vpt 31.5 def
/M {moveto} bind def
/L {lineto} bind def
/R {rmoveto} bind def
/V {rlineto} bind def
/vpt2 vpt 2 mul def
/hpt2 hpt 2 mul def
/Lshow { currentpoint stroke M
  0 vshift R show } def
/Rshow { currentpoint stroke M
  dup stringwidth pop neg vshift R show } def
/Cshow { currentpoint stroke M
  dup stringwidth pop -2 div vshift R show } def
/DL { Color {setrgbcolor Solid {pop []} if 0 setdash }
 {pop pop pop Solid {pop []} if 0 setdash} ifelse } def
/BL { stroke gnulinewidth 2 mul setlinewidth } def
/AL { stroke gnulinewidth 2 div setlinewidth } def
/PL { stroke gnulinewidth setlinewidth } def
/LTb { BL [] 0 0 0 DL } def
/LTa { AL [1 dl 2 dl] 0 setdash 0 0 0 setrgbcolor } def
/LT0 { PL [] 0 1 0 DL } def
/LT1 { PL [4 dl 2 dl] 0 0 1 DL } def
/LT2 { PL [2 dl 3 dl] 1 0 0 DL } def
/LT3 { PL [1 dl 1.5 dl] 1 0 1 DL } def
/LT4 { PL [5 dl 2 dl 1 dl 2 dl] 0 1 1 DL } def
/LT5 { PL [4 dl 3 dl 1 dl 3 dl] 1 1 0 DL } def
/LT6 { PL [2 dl 2 dl 2 dl 4 dl] 0 0 0 DL } def
/LT7 { PL [2 dl 2 dl 2 dl 2 dl 2 dl 4 dl] 1 0.3 0 DL } def
/LT8 { PL [2 dl 2 dl 2 dl 2 dl 2 dl 2 dl 2 dl 4 dl] 0.5 0.5 0.5 DL } def
/P { stroke [] 0 setdash
  currentlinewidth 2 div sub M
  0 currentlinewidth V stroke } def
/D { stroke [] 0 setdash 2 copy vpt add M
  hpt neg vpt neg V hpt vpt neg V
  hpt vpt V hpt neg vpt V closepath stroke
  P } def
/A { stroke [] 0 setdash vpt sub M 0 vpt2 V
  currentpoint stroke M
  hpt neg vpt neg R hpt2 0 V stroke
  } def
/B { stroke [] 0 setdash 2 copy exch hpt sub exch vpt add M
  0 vpt2 neg V hpt2 0 V 0 vpt2 V
  hpt2 neg 0 V closepath stroke
  P } def
/C { stroke [] 0 setdash exch hpt sub exch vpt add M
  hpt2 vpt2 neg V currentpoint stroke M
  hpt2 neg 0 R hpt2 vpt2 V stroke } def
/T { stroke [] 0 setdash 2 copy vpt 1.12 mul add M
  hpt neg vpt -1.62 mul V
  hpt 2 mul 0 V
  hpt neg vpt 1.62 mul V closepath stroke
  P  } def
/S { 2 copy A C} def
end
}
\begin{picture}(2519,1511)(0,0)
\special{"
gnudict begin
gsave
50 50 translate
0.100 0.100 scale
0 setgray
/Helvetica findfont 100 scalefont setfont
newpath
-500.000000 -500.000000 translate
LTa
600 251 M
1736 0 V
-868 0 R
0 1209 V
LTb
600 251 M
63 0 V
1673 0 R
-63 0 V
600 424 M
63 0 V
1673 0 R
-63 0 V
600 596 M
63 0 V
1673 0 R
-63 0 V
600 769 M
63 0 V
1673 0 R
-63 0 V
600 942 M
63 0 V
1673 0 R
-63 0 V
600 1115 M
63 0 V
1673 0 R
-63 0 V
600 1287 M
63 0 V
1673 0 R
-63 0 V
600 1460 M
63 0 V
1673 0 R
-63 0 V
600 251 M
0 63 V
0 1146 R
0 -63 V
817 251 M
0 63 V
0 1146 R
0 -63 V
1034 251 M
0 63 V
0 1146 R
0 -63 V
1251 251 M
0 63 V
0 1146 R
0 -63 V
1468 251 M
0 63 V
0 1146 R
0 -63 V
1685 251 M
0 63 V
0 1146 R
0 -63 V
1902 251 M
0 63 V
0 1146 R
0 -63 V
2119 251 M
0 63 V
0 1146 R
0 -63 V
2336 251 M
0 63 V
0 1146 R
0 -63 V
600 251 M
1736 0 V
0 1209 V
-1736 0 V
600 251 L
LT0
618 1156 M
17 20 V
18 15 V
17 12 V
18 10 V
17 9 V
18 9 V
17 7 V
18 7 V
17 6 V
18 6 V
17 5 V
18 5 V
17 5 V
18 4 V
18 4 V
17 4 V
18 4 V
17 4 V
18 3 V
17 3 V
18 3 V
17 3 V
18 2 V
17 3 V
18 2 V
17 2 V
18 3 V
18 2 V
17 1 V
18 2 V
17 2 V
18 1 V
17 2 V
18 1 V
17 1 V
18 2 V
17 1 V
18 1 V
17 0 V
18 1 V
17 1 V
18 0 V
18 1 V
17 0 V
18 1 V
17 0 V
18 0 V
17 0 V
18 0 V
17 0 V
18 0 V
17 0 V
18 -1 V
17 0 V
18 -1 V
18 0 V
17 -1 V
18 -1 V
17 0 V
18 -1 V
17 -1 V
18 -2 V
17 -1 V
18 -1 V
17 -2 V
18 -1 V
17 -2 V
18 -2 V
17 -1 V
18 -2 V
18 -3 V
17 -2 V
18 -2 V
17 -3 V
18 -2 V
17 -3 V
18 -3 V
17 -3 V
18 -3 V
17 -4 V
18 -4 V
17 -4 V
18 -4 V
18 -4 V
17 -5 V
18 -5 V
17 -5 V
18 -6 V
17 -6 V
18 -7 V
17 -7 V
18 -9 V
17 -9 V
18 -10 V
17 -12 V
18 -15 V
17 -20 V
LT1
1685 933 D
1468 1029 D
1251 1126 D
1034 1209 D
947 1231 D
904 1238 D
860 1243 D
852 1243 D
817 1243 D
774 1239 D
730 1229 D
687 1212 D
643 1183 D
LT2
1685 627 A
1468 793 A
1251 990 A
1034 1168 A
947 1215 A
904 1231 A
860 1241 A
839 1244 A
817 1245 A
774 1241 A
730 1231 A
687 1212 A
643 1183 A
622 1161 A
LT3
1685 339 B
1468 413 B
1251 567 B
1034 839 B
947 969 B
904 1033 B
860 1090 B
856 1096 B
852 1101 B
839 1116 B
817 1139 B
774 1176 B
730 1196 B
687 1198 B
643 1177 B
622 1157 B
600 1116 B
LT4
1685 251 C
1468 269 C
1251 343 C
1034 662 C
947 876 C
904 981 C
860 1072 C
856 1081 C
852 1088 C
839 1111 C
774 1188 C
730 1207 C
817 1143 C
687 1202 C
643 1178 C
stroke
grestore
end
showpage
}
\put(665,545){\makebox(0,0)[l]{$0.05$, $100$ }}
\put(1511,424){\makebox(0,0)[l]{$0.05$, $50$ }}
\put(1520,804){\makebox(0,0)[l]{$0.01$, $100$ }}
\put(1316,1132){\makebox(0,0)[l]{$0.01$, $50$ }}
\put(1468,51){\makebox(0,0){$V$}}
\put(100,855){%
\special{ps: gsave currentpoint currentpoint translate
270 rotate neg exch neg exch translate}%
\makebox(0,0)[b]{\shortstack{$M\Delta E(u)$}}%
\special{ps: currentpoint grestore moveto}%
}
\put(2336,151){\makebox(0,0){2}}
\put(2119,151){\makebox(0,0){1.5}}
\put(1902,151){\makebox(0,0){1}}
\put(1685,151){\makebox(0,0){0.5}}
\put(1468,151){\makebox(0,0){0}}
\put(1251,151){\makebox(0,0){-0.5}}
\put(1034,151){\makebox(0,0){-1}}
\put(817,151){\makebox(0,0){-1.5}}
\put(600,151){\makebox(0,0){-2}}
\put(540,1460){\makebox(0,0)[r]{3.5}}
\put(540,1287){\makebox(0,0)[r]{3}}
\put(540,1115){\makebox(0,0)[r]{2.5}}
\put(540,942){\makebox(0,0)[r]{2}}
\put(540,769){\makebox(0,0)[r]{1.5}}
\put(540,596){\makebox(0,0)[r]{1}}
\put(540,424){\makebox(0,0)[r]{0.5}}
\put(540,251){\makebox(0,0)[r]{0}}
\end{picture}
 FIG. \ref{fig3}: \refstepcounter{figure} \label{fig3}
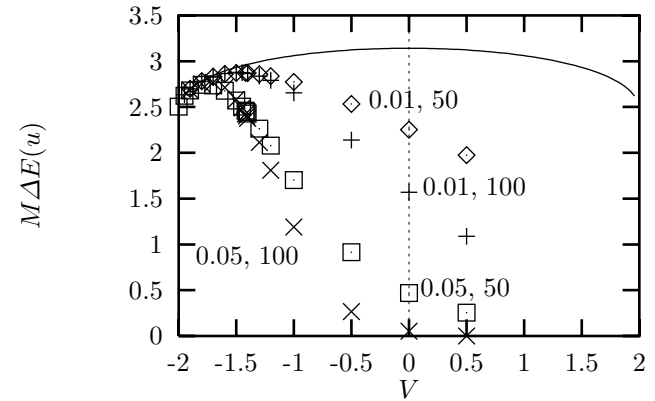
Phase sensitivity versus interaction,
for system sizes 50 and 100; $u = 0.01$ and 0.05, $\lambda_t=1$, $\lambda_V=0$.
For comparison, we include here and in figures 4--6, the clean 
limit result (full line).
 \\ \\  
\setlength{\unitlength}{0.1bp}
\special{!
/gnudict 40 dict def
gnudict begin
/Color false def
/Solid false def
/gnulinewidth 5.000 def
/vshift -33 def
/dl {10 mul} def
/hpt 31.5 def
/vpt 31.5 def
/M {moveto} bind def
/L {lineto} bind def
/R {rmoveto} bind def
/V {rlineto} bind def
/vpt2 vpt 2 mul def
/hpt2 hpt 2 mul def
/Lshow { currentpoint stroke M
  0 vshift R show } def
/Rshow { currentpoint stroke M
  dup stringwidth pop neg vshift R show } def
/Cshow { currentpoint stroke M
  dup stringwidth pop -2 div vshift R show } def
/DL { Color {setrgbcolor Solid {pop []} if 0 setdash }
 {pop pop pop Solid {pop []} if 0 setdash} ifelse } def
/BL { stroke gnulinewidth 2 mul setlinewidth } def
/AL { stroke gnulinewidth 2 div setlinewidth } def
/PL { stroke gnulinewidth setlinewidth } def
/LTb { BL [] 0 0 0 DL } def
/LTa { AL [1 dl 2 dl] 0 setdash 0 0 0 setrgbcolor } def
/LT0 { PL [] 0 1 0 DL } def
/LT1 { PL [4 dl 2 dl] 0 0 1 DL } def
/LT2 { PL [2 dl 3 dl] 1 0 0 DL } def
/LT3 { PL [1 dl 1.5 dl] 1 0 1 DL } def
/LT4 { PL [5 dl 2 dl 1 dl 2 dl] 0 1 1 DL } def
/LT5 { PL [4 dl 3 dl 1 dl 3 dl] 1 1 0 DL } def
/LT6 { PL [2 dl 2 dl 2 dl 4 dl] 0 0 0 DL } def
/LT7 { PL [2 dl 2 dl 2 dl 2 dl 2 dl 4 dl] 1 0.3 0 DL } def
/LT8 { PL [2 dl 2 dl 2 dl 2 dl 2 dl 2 dl 2 dl 4 dl] 0.5 0.5 0.5 DL } def
/P { stroke [] 0 setdash
  currentlinewidth 2 div sub M
  0 currentlinewidth V stroke } def
/D { stroke [] 0 setdash 2 copy vpt add M
  hpt neg vpt neg V hpt vpt neg V
  hpt vpt V hpt neg vpt V closepath stroke
  P } def
/A { stroke [] 0 setdash vpt sub M 0 vpt2 V
  currentpoint stroke M
  hpt neg vpt neg R hpt2 0 V stroke
  } def
/B { stroke [] 0 setdash 2 copy exch hpt sub exch vpt add M
  0 vpt2 neg V hpt2 0 V 0 vpt2 V
  hpt2 neg 0 V closepath stroke
  P } def
/C { stroke [] 0 setdash exch hpt sub exch vpt add M
  hpt2 vpt2 neg V currentpoint stroke M
  hpt2 neg 0 R hpt2 vpt2 V stroke } def
/T { stroke [] 0 setdash 2 copy vpt 1.12 mul add M
  hpt neg vpt -1.62 mul V
  hpt 2 mul 0 V
  hpt neg vpt 1.62 mul V closepath stroke
  P  } def
/S { 2 copy A C} def
end
}
\begin{picture}(2519,1511)(0,0)
\special{"
gnudict begin
gsave
50 50 translate
0.100 0.100 scale
0 setgray
/Helvetica findfont 100 scalefont setfont
newpath
-500.000000 -500.000000 translate
LTa
600 251 M
1736 0 V
LTb
600 251 M
63 0 V
1673 0 R
-63 0 V
600 424 M
63 0 V
1673 0 R
-63 0 V
600 596 M
63 0 V
1673 0 R
-63 0 V
600 769 M
63 0 V
1673 0 R
-63 0 V
600 942 M
63 0 V
1673 0 R
-63 0 V
600 1115 M
63 0 V
1673 0 R
-63 0 V
600 1287 M
63 0 V
1673 0 R
-63 0 V
600 1460 M
63 0 V
1673 0 R
-63 0 V
600 251 M
0 63 V
0 1146 R
0 -63 V
1034 251 M
0 63 V
0 1146 R
0 -63 V
1468 251 M
0 63 V
0 1146 R
0 -63 V
1902 251 M
0 63 V
0 1146 R
0 -63 V
2336 251 M
0 63 V
0 1146 R
0 -63 V
600 251 M
1736 0 V
0 1209 V
-1736 0 V
600 251 L
LT0
618 1141 M
17 15 V
18 11 V
17 9 V
18 8 V
17 7 V
18 6 V
17 6 V
18 5 V
17 5 V
18 5 V
17 4 V
18 5 V
17 4 V
18 3 V
18 4 V
17 3 V
18 4 V
17 3 V
18 3 V
17 3 V
18 3 V
17 3 V
18 2 V
17 3 V
18 2 V
17 3 V
18 2 V
18 2 V
17 2 V
18 2 V
17 2 V
18 2 V
17 2 V
18 2 V
17 2 V
18 2 V
17 2 V
18 1 V
17 2 V
18 1 V
17 2 V
18 2 V
18 1 V
17 1 V
18 2 V
17 1 V
18 1 V
17 2 V
18 1 V
17 1 V
18 1 V
17 1 V
18 1 V
17 2 V
18 1 V
18 1 V
17 1 V
18 1 V
17 0 V
18 1 V
17 1 V
18 1 V
17 1 V
18 1 V
17 0 V
18 1 V
17 1 V
18 0 V
17 1 V
18 1 V
18 0 V
17 1 V
18 1 V
17 0 V
18 1 V
17 0 V
18 1 V
17 0 V
18 0 V
17 1 V
18 0 V
17 0 V
18 1 V
18 0 V
17 0 V
18 1 V
17 0 V
18 0 V
17 0 V
18 1 V
17 0 V
18 0 V
17 0 V
18 0 V
17 0 V
18 0 V
17 0 V
18 0 V
LT1
2109 1184 D
2336 269 D
1902 310 D
1468 482 D
1294 639 D
1208 738 D
1121 846 D
1077 251 D
1034 953 D
947 1049 D
860 1121 D
774 1159 D
730 1163 D
687 1158 D
643 1143 D
600 251 D
LT2
2109 1084 A
2336 251 A
1902 253 A
1468 313 A
1294 459 A
1208 595 A
1121 761 A
1077 846 A
1034 927 A
947 1061 A
860 1142 A
774 1170 A
730 1169 A
687 1159 A
643 1143 A
600 1110 A
LT3
2109 984 B
1121 353 B
1034 440 B
947 579 B
860 758 B
774 932 B
730 999 B
687 1044 B
643 1065 B
LT4
2109 884 C
1121 265 C
1034 311 C
947 458 C
860 723 C
817 855 C
774 960 C
730 1028 C
687 1063 C
643 1072 C
600 1065 C
LT5
2109 784 T
1121 253 T
1034 257 T
947 267 T
860 299 T
774 396 T
730 490 T
687 612 T
643 745 T
LT6
2109 684 S
1121 251 S
1034 251 S
947 251 S
860 255 S
817 265 S
774 299 S
730 396 S
687 575 S
643 766 S
600 893 S
LT7
2109 584 D
860 251 D
774 252 D
730 254 D
687 257 D
643 266 D
600 290 D
LT8
2109 484 A
947 251 A
860 251 A
817 251 A
774 251 A
730 251 A
687 251 A
643 251 A
600 330 A
600 255 A
stroke
grestore
end
showpage
}
\put(1989,484){\makebox(0,0)[r]{}}
\put(1989,534){\makebox(0,0)[r]{$0.4$}}
\put(1989,684){\makebox(0,0)[r]{}}
\put(1989,734){\makebox(0,0)[r]{$0.3$}}
\put(1989,884){\makebox(0,0)[r]{}}
\put(1989,934){\makebox(0,0)[r]{$0.2$}}
\put(1989,1084){\makebox(0,0)[r]{}}
\put(1989,1134){\makebox(0,0)[r]{$0.1$}}
\put(1468,51){\makebox(0,0){$V$}}
\put(100,855){%
\special{ps: gsave currentpoint currentpoint translate
270 rotate neg exch neg exch translate}%
\makebox(0,0)[b]{\shortstack{$M\Delta E(u)$}}%
\special{ps: currentpoint grestore moveto}%
}
\put(2336,151){\makebox(0,0){0}}
\put(1902,151){\makebox(0,0){-0.5}}
\put(1468,151){\makebox(0,0){-1}}
\put(1034,151){\makebox(0,0){-1.5}}
\put(600,151){\makebox(0,0){-2}}
\put(540,1460){\makebox(0,0)[r]{3.5}}
\put(540,1287){\makebox(0,0)[r]{3}}
\put(540,1115){\makebox(0,0)[r]{2.5}}
\put(540,942){\makebox(0,0)[r]{2}}
\put(540,769){\makebox(0,0)[r]{1.5}}
\put(540,596){\makebox(0,0)[r]{1}}
\put(540,424){\makebox(0,0)[r]{0.5}}
\put(540,251){\makebox(0,0)[r]{0}}
\end{picture}
FIG. \ref{fig4}: \refstepcounter{figure} \label{fig4}
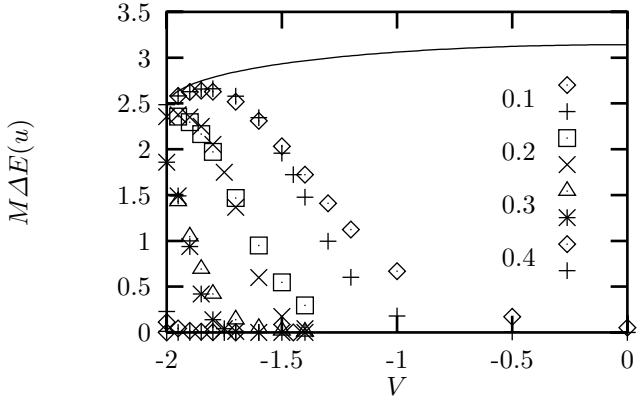
Phase sensitivity versus interaction, for 
 $u = 0.1 \ldots 0.4$; $\lambda_t=1$, $ \lambda_V=0$.
 The respective upper data correspond to $M=50$, the lower ones to $M=100$.
  \\ \\ 
\setlength{\unitlength}{0.1bp}
\special{!
/gnudict 40 dict def
gnudict begin
/Color false def
/Solid false def
/gnulinewidth 5.000 def
/vshift -33 def
/dl {10 mul} def
/hpt 31.5 def
/vpt 31.5 def
/M {moveto} bind def
/L {lineto} bind def
/R {rmoveto} bind def
/V {rlineto} bind def
/vpt2 vpt 2 mul def
/hpt2 hpt 2 mul def
/Lshow { currentpoint stroke M
  0 vshift R show } def
/Rshow { currentpoint stroke M
  dup stringwidth pop neg vshift R show } def
/Cshow { currentpoint stroke M
  dup stringwidth pop -2 div vshift R show } def
/DL { Color {setrgbcolor Solid {pop []} if 0 setdash }
 {pop pop pop Solid {pop []} if 0 setdash} ifelse } def
/BL { stroke gnulinewidth 2 mul setlinewidth } def
/AL { stroke gnulinewidth 2 div setlinewidth } def
/PL { stroke gnulinewidth setlinewidth } def
/LTb { BL [] 0 0 0 DL } def
/LTa { AL [1 dl 2 dl] 0 setdash 0 0 0 setrgbcolor } def
/LT0 { PL [] 0 1 0 DL } def
/LT1 { PL [4 dl 2 dl] 0 0 1 DL } def
/LT2 { PL [2 dl 3 dl] 1 0 0 DL } def
/LT3 { PL [1 dl 1.5 dl] 1 0 1 DL } def
/LT4 { PL [5 dl 2 dl 1 dl 2 dl] 0 1 1 DL } def
/LT5 { PL [4 dl 3 dl 1 dl 3 dl] 1 1 0 DL } def
/LT6 { PL [2 dl 2 dl 2 dl 4 dl] 0 0 0 DL } def
/LT7 { PL [2 dl 2 dl 2 dl 2 dl 2 dl 4 dl] 1 0.3 0 DL } def
/LT8 { PL [2 dl 2 dl 2 dl 2 dl 2 dl 2 dl 2 dl 4 dl] 0.5 0.5 0.5 DL } def
/P { stroke [] 0 setdash
  currentlinewidth 2 div sub M
  0 currentlinewidth V stroke } def
/D { stroke [] 0 setdash 2 copy vpt add M
  hpt neg vpt neg V hpt vpt neg V
  hpt vpt V hpt neg vpt V closepath stroke
  P } def
/A { stroke [] 0 setdash vpt sub M 0 vpt2 V
  currentpoint stroke M
  hpt neg vpt neg R hpt2 0 V stroke
  } def
/B { stroke [] 0 setdash 2 copy exch hpt sub exch vpt add M
  0 vpt2 neg V hpt2 0 V 0 vpt2 V
  hpt2 neg 0 V closepath stroke
  P } def
/C { stroke [] 0 setdash exch hpt sub exch vpt add M
  hpt2 vpt2 neg V currentpoint stroke M
  hpt2 neg 0 R hpt2 vpt2 V stroke } def
/T { stroke [] 0 setdash 2 copy vpt 1.12 mul add M
  hpt neg vpt -1.62 mul V
  hpt 2 mul 0 V
  hpt neg vpt 1.62 mul V closepath stroke
  P  } def
/S { 2 copy A C} def
end
}
\begin{picture}(2519,1511)(0,0)
\special{"
gnudict begin
gsave
50 50 translate
0.100 0.100 scale
0 setgray
/Helvetica findfont 100 scalefont setfont
newpath
-500.000000 -500.000000 translate
LTa
600 251 M
1736 0 V
-868 0 R
0 1209 V
LTb
600 251 M
63 0 V
1673 0 R
-63 0 V
600 424 M
63 0 V
1673 0 R
-63 0 V
600 596 M
63 0 V
1673 0 R
-63 0 V
600 769 M
63 0 V
1673 0 R
-63 0 V
600 942 M
63 0 V
1673 0 R
-63 0 V
600 1115 M
63 0 V
1673 0 R
-63 0 V
600 1287 M
63 0 V
1673 0 R
-63 0 V
600 1460 M
63 0 V
1673 0 R
-63 0 V
600 251 M
0 63 V
0 1146 R
0 -63 V
817 251 M
0 63 V
0 1146 R
0 -63 V
1034 251 M
0 63 V
0 1146 R
0 -63 V
1251 251 M
0 63 V
0 1146 R
0 -63 V
1468 251 M
0 63 V
0 1146 R
0 -63 V
1685 251 M
0 63 V
0 1146 R
0 -63 V
1902 251 M
0 63 V
0 1146 R
0 -63 V
2119 251 M
0 63 V
0 1146 R
0 -63 V
2336 251 M
0 63 V
0 1146 R
0 -63 V
600 251 M
1736 0 V
0 1209 V
-1736 0 V
600 251 L
LT0
618 1156 M
17 20 V
18 15 V
17 12 V
18 10 V
17 9 V
18 9 V
17 7 V
18 7 V
17 6 V
18 6 V
17 5 V
18 5 V
17 5 V
18 4 V
18 4 V
17 4 V
18 4 V
17 4 V
18 3 V
17 3 V
18 3 V
17 3 V
18 2 V
17 3 V
18 2 V
17 2 V
18 3 V
18 2 V
17 1 V
18 2 V
17 2 V
18 1 V
17 2 V
18 1 V
17 1 V
18 2 V
17 1 V
18 1 V
17 0 V
18 1 V
17 1 V
18 0 V
18 1 V
17 0 V
18 1 V
17 0 V
18 0 V
17 0 V
18 0 V
17 0 V
18 0 V
17 0 V
18 -1 V
17 0 V
18 -1 V
18 0 V
17 -1 V
18 -1 V
17 0 V
18 -1 V
17 -1 V
18 -2 V
17 -1 V
18 -1 V
17 -2 V
18 -1 V
17 -2 V
18 -2 V
17 -1 V
18 -2 V
18 -3 V
17 -2 V
18 -2 V
17 -3 V
18 -2 V
17 -3 V
18 -3 V
17 -3 V
18 -3 V
17 -4 V
18 -4 V
17 -4 V
18 -4 V
18 -4 V
17 -5 V
18 -5 V
17 -5 V
18 -6 V
17 -6 V
18 -7 V
17 -7 V
18 -9 V
17 -9 V
18 -10 V
17 -12 V
18 -15 V
17 -20 V
LT1
2217 1149 D
2336 570 D
1902 755 D
1685 886 D
1468 1029 D
1251 1159 D
1034 1246 D
947 1261 D
904 1263 D
860 1262 D
856 1262 D
852 1261 D
839 1260 D
817 1257 D
730 1233 D
687 1213 D
643 1184 D
622 1162 D
LT2
2217 1049 A
1902 422 A
1685 569 A
1468 793 A
1251 1043 A
1034 1220 A
947 1253 A
904 1260 A
860 1262 A
856 1261 A
852 1261 A
839 1260 A
817 1257 A
730 1234 A
687 1213 A
643 1183 A
622 1161 A
600 1114 A
LT3
2217 949 B
2336 256 B
1902 274 B
1468 413 B
1251 654 B
1034 1005 B
947 1123 B
904 1168 B
860 1201 B
856 1204 B
852 1206 B
839 1213 B
817 1222 B
730 1225 B
687 1210 B
643 1181 B
622 1160 B
LT4
2217 849 C
1902 251 C
1685 253 C
1468 269 C
1251 405 C
1034 884 C
947 1079 C
904 1149 C
860 1198 C
856 1201 C
852 1205 C
839 1247 C
817 1225 C
730 1228 C
687 1210 C
643 1181 C
600 1112 C
LT5
2217 749 T
1034 727 T
947 939 T
860 1111 T
817 1167 T
774 1200 T
730 1210 T
687 1201 T
643 1174 T
622 1153 T
LT6
2217 649 S
1034 529 S
947 838 S
860 1098 S
817 1171 S
774 1208 S
730 1215 S
687 1202 S
643 1174 S
622 1152 S
stroke
grestore
end
showpage
}
\put(2097,699){\makebox(0,0)[r]{$0.1$}}
\put(2097,749){\makebox(0,0)[r]{}}
\put(2097,899){\makebox(0,0)[r]{$0.05$}}
\put(2097,949){\makebox(0,0)[r]{}}
\put(2097,1099){\makebox(0,0)[r]{$0.01$}}
\put(2097,1149){\makebox(0,0)[r]{}}
\put(1468,51){\makebox(0,0){$V$}}
\put(100,855){%
\special{ps: gsave currentpoint currentpoint translate
270 rotate neg exch neg exch translate}%
\makebox(0,0)[b]{\shortstack{$M\Delta E(u)$}}%
\special{ps: currentpoint grestore moveto}%
}
\put(2336,151){\makebox(0,0){2}}
\put(2119,151){\makebox(0,0){1.5}}
\put(1902,151){\makebox(0,0){1}}
\put(1685,151){\makebox(0,0){0.5}}
\put(1468,151){\makebox(0,0){0}}
\put(1251,151){\makebox(0,0){-0.5}}
\put(1034,151){\makebox(0,0){-1}}
\put(817,151){\makebox(0,0){-1.5}}
\put(600,151){\makebox(0,0){-2}}
\put(540,1460){\makebox(0,0)[r]{3.5}}
\put(540,1287){\makebox(0,0)[r]{3}}
\put(540,1115){\makebox(0,0)[r]{2.5}}
\put(540,942){\makebox(0,0)[r]{2}}
\put(540,769){\makebox(0,0)[r]{1.5}}
\put(540,596){\makebox(0,0)[r]{1}}
\put(540,424){\makebox(0,0)[r]{0.5}}
\put(540,251){\makebox(0,0)[r]{0}}
\end{picture}
 FIG. \ref{fig5}: \refstepcounter{figure} \label{fig5}
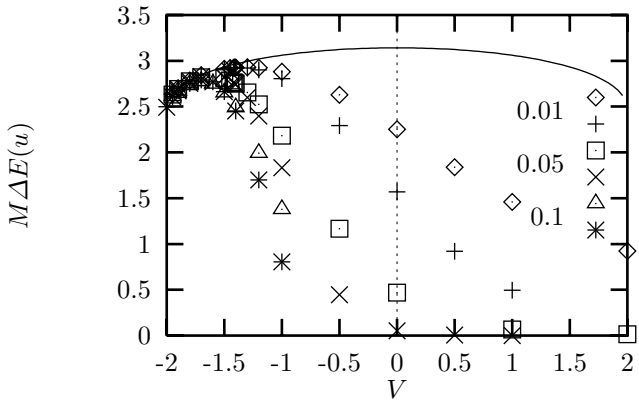
Phase sensitivity versus  interaction, for 
$u= 0.01 \ldots 0.1$; $\lambda_V=\lambda_t=1$. 
The respective upper data correspond to $M=50$, the lower ones to $M=100$.
\\ \\  
\setlength{\unitlength}{0.1bp}
\begin{picture}(2519,1511)(0,0)
\put(1989,590){\makebox(0,0)[r]{}}
\put(1989,640){\makebox(0,0)[r]{$0.8$}}
\put(1989,790){\makebox(0,0)[r]{}}
\put(1989,840){\makebox(0,0)[r]{$0.6$}}
\put(1989,990){\makebox(0,0)[r]{}}
\put(1989,1040){\makebox(0,0)[r]{$0.4$}}
\put(1989,1190){\makebox(0,0)[r]{}}
\put(1989,1240){\makebox(0,0)[r]{$0.2$}}
\put(1468,51){\makebox(0,0){$V$}}
\put(100,855){%
\makebox(0,0)[b]{\shortstack{$M\Delta E(u)$}}%
}
\put(2336,151){\makebox(0,0){0}}
\put(1902,151){\makebox(0,0){-0.5}}
\put(1468,151){\makebox(0,0){-1}}
\put(1034,151){\makebox(0,0){-1.5}}
\put(600,151){\makebox(0,0){-2}}
\put(540,1406){\makebox(0,0)[r]{3}}
\put(540,1213){\makebox(0,0)[r]{2.5}}
\put(540,1021){\makebox(0,0)[r]{2}}
\put(540,828){\makebox(0,0)[r]{1.5}}
\put(540,636){\makebox(0,0)[r]{1}}
\put(540,443){\makebox(0,0)[r]{0.5}}
\put(540,251){\makebox(0,0)[r]{0}}
\end{picture}

 FIG. \ref{fig6}: \refstepcounter{figure} \label{fig6}
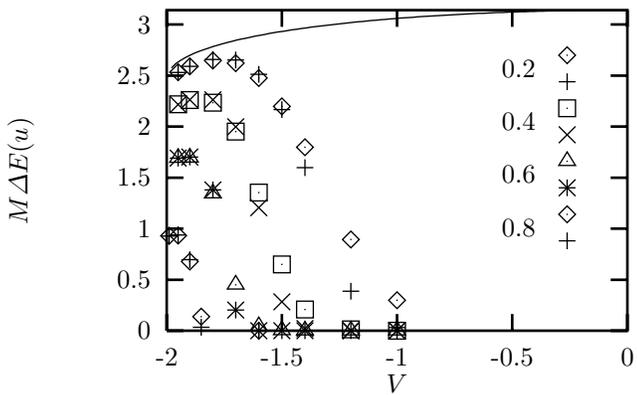
Phase sensitivity versus interaction, for
$ u = 0.2\ldots  0.8$; $\lambda_V=\lambda_t=1$. 
The respective upper data correspond to $M=50$, and the lower ones to $M=100$.
\\ \\  
\setlength{\unitlength}{0.1bp}
\begin{picture}(2519,1511)(0,0)
\put(2052,573){\makebox(0,0)[l]{$V=1$}}
\put(1926,936){\makebox(0,0)[l]{$V=0$}}
\put(1957,1158){\makebox(0,0)[l]{$V=-1$}}
\put(1452,654){\makebox(0,0)[l]{$V=2$}}
\put(2178,1359){\makebox(0,0)[r]{$V=-1.9$}}
\put(758,453){\makebox(0,0)[l]{$M=200$, 350 states}}
\put(1468,51){\makebox(0,0){$\lambda u$}}
\put(100,855){%
\makebox(0,0)[b]{\shortstack{$\Delta E( \lambda u)$}}%
}
\put(2178,151){\makebox(0,0){0.05}}
\put(1863,151){\makebox(0,0){0.04}}
\put(1547,151){\makebox(0,0){0.03}}
\put(1231,151){\makebox(0,0){0.02}}
\put(916,151){\makebox(0,0){0.01}}
\put(600,151){\makebox(0,0){0}}
\put(540,1460){\makebox(0,0)[r]{0.5}}
\put(540,1259){\makebox(0,0)[r]{0}}
\put(540,1057){\makebox(0,0)[r]{-0.5}}
\put(540,856){\makebox(0,0)[r]{-1}}
\put(540,654){\makebox(0,0)[r]{-1.5}}
\put(540,453){\makebox(0,0)[r]{-2}}
\put(540,251){\makebox(0,0)[r]{-2.5}}
\end{picture}

 FIG. \ref{fig7}: \refstepcounter{figure} \label{fig7}
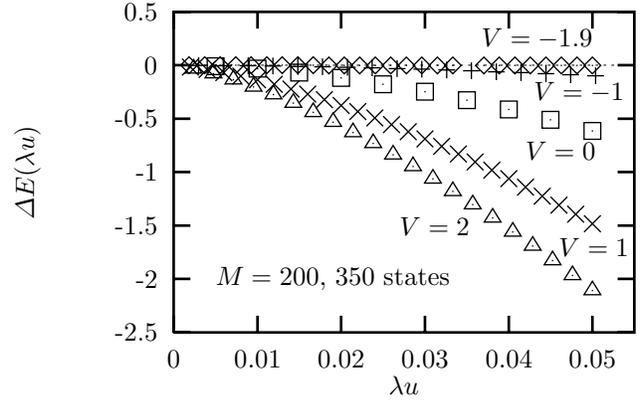
Ground state energy versus $\lambda u $.  The  system size is  $M=200$.
 \\ \\ 
\setlength{\unitlength}{0.1bp}
\begin{picture}(2519,1511)(0,0)
\put(1034,556){\makebox(0,0)[r]{$V=4.0$}}
\put(1034,656){\makebox(0,0)[r]{$V=3.5$}}
\put(1034,756){\makebox(0,0)[r]{$V=3.0$}}
\put(1034,856){\makebox(0,0)[r]{$V=2.1$}}
\put(1468,51){\makebox(0,0){$\lambda u$  }}
\put(100,855){%
\makebox(0,0)[b]{\shortstack{$\Delta E(\lambda u)$}}%
}
\put(2336,151){\makebox(0,0){0.08}}
\put(1902,151){\makebox(0,0){0.06}}
\put(1468,151){\makebox(0,0){0.04}}
\put(1034,151){\makebox(0,0){0.02}}
\put(600,151){\makebox(0,0){0}}
\put(540,1460){\makebox(0,0)[r]{0.5}}
\put(540,1259){\makebox(0,0)[r]{0}}
\put(540,1057){\makebox(0,0)[r]{-0.5}}
\put(540,856){\makebox(0,0)[r]{-1}}
\put(540,654){\makebox(0,0)[r]{-1.5}}
\put(540,453){\makebox(0,0)[r]{-2}}
\put(540,251){\makebox(0,0)[r]{-2.5}}
\end{picture}

FIG. \ref{fig8}: \refstepcounter{figure} \label{fig8}
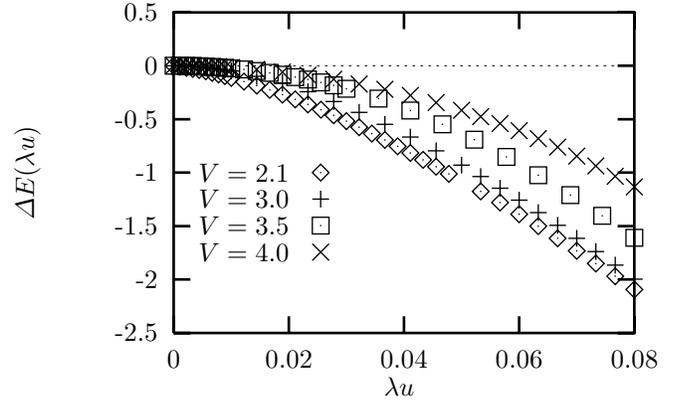
Ground state energy versus $\lambda u$. The system size is $M=100$.
\\  \\ 
\setlength{\unitlength}{0.1bp}
\special{!
/gnudict 40 dict def
gnudict begin
/Color false def
/Solid false def
/gnulinewidth 5.000 def
/vshift -33 def
/dl {10 mul} def
/hpt 31.5 def
/vpt 31.5 def
/M {moveto} bind def
/L {lineto} bind def
/R {rmoveto} bind def
/V {rlineto} bind def
/vpt2 vpt 2 mul def
/hpt2 hpt 2 mul def
/Lshow { currentpoint stroke M
  0 vshift R show } def
/Rshow { currentpoint stroke M
  dup stringwidth pop neg vshift R show } def
/Cshow { currentpoint stroke M
  dup stringwidth pop -2 div vshift R show } def
/DL { Color {setrgbcolor Solid {pop []} if 0 setdash }
 {pop pop pop Solid {pop []} if 0 setdash} ifelse } def
/BL { stroke gnulinewidth 2 mul setlinewidth } def
/AL { stroke gnulinewidth 2 div setlinewidth } def
/PL { stroke gnulinewidth setlinewidth } def
/LTb { BL [] 0 0 0 DL } def
/LTa { AL [1 dl 2 dl] 0 setdash 0 0 0 setrgbcolor } def
/LT0 { PL [] 0 1 0 DL } def
/LT1 { PL [4 dl 2 dl] 0 0 1 DL } def
/LT2 { PL [2 dl 3 dl] 1 0 0 DL } def
/LT3 { PL [1 dl 1.5 dl] 1 0 1 DL } def
/LT4 { PL [5 dl 2 dl 1 dl 2 dl] 0 1 1 DL } def
/LT5 { PL [4 dl 3 dl 1 dl 3 dl] 1 1 0 DL } def
/LT6 { PL [2 dl 2 dl 2 dl 4 dl] 0 0 0 DL } def
/LT7 { PL [2 dl 2 dl 2 dl 2 dl 2 dl 4 dl] 1 0.3 0 DL } def
/LT8 { PL [2 dl 2 dl 2 dl 2 dl 2 dl 2 dl 2 dl 4 dl] 0.5 0.5 0.5 DL } def
/P { stroke [] 0 setdash
  currentlinewidth 2 div sub M
  0 currentlinewidth V stroke } def
/D { stroke [] 0 setdash 2 copy vpt add M
  hpt neg vpt neg V hpt vpt neg V
  hpt vpt V hpt neg vpt V closepath stroke
  P } def
/A { stroke [] 0 setdash vpt sub M 0 vpt2 V
  currentpoint stroke M
  hpt neg vpt neg R hpt2 0 V stroke
  } def
/B { stroke [] 0 setdash 2 copy exch hpt sub exch vpt add M
  0 vpt2 neg V hpt2 0 V 0 vpt2 V
  hpt2 neg 0 V closepath stroke
  P } def
/C { stroke [] 0 setdash exch hpt sub exch vpt add M
  hpt2 vpt2 neg V currentpoint stroke M
  hpt2 neg 0 R hpt2 vpt2 V stroke } def
/T { stroke [] 0 setdash 2 copy vpt 1.12 mul add M
  hpt neg vpt -1.62 mul V
  hpt 2 mul 0 V
  hpt neg vpt 1.62 mul V closepath stroke
  P  } def
/S { 2 copy A C} def
end
}
\begin{picture}(2519,1511)(0,0)
\special{"
gnudict begin
gsave
50 50 translate
0.100 0.100 scale
0 setgray
/Helvetica findfont 100 scalefont setfont
newpath
-500.000000 -500.000000 translate
LTa
600 251 M
0 1209 V
LTb
600 251 M
63 0 V
1673 0 R
-63 0 V
600 493 M
63 0 V
1673 0 R
-63 0 V
600 735 M
63 0 V
1673 0 R
-63 0 V
600 976 M
63 0 V
1673 0 R
-63 0 V
600 1218 M
63 0 V
1673 0 R
-63 0 V
600 1460 M
63 0 V
1673 0 R
-63 0 V
600 251 M
0 63 V
0 1146 R
0 -63 V
774 251 M
0 63 V
0 1146 R
0 -63 V
947 251 M
0 63 V
0 1146 R
0 -63 V
1121 251 M
0 63 V
0 1146 R
0 -63 V
1294 251 M
0 63 V
0 1146 R
0 -63 V
1468 251 M
0 63 V
0 1146 R
0 -63 V
1642 251 M
0 63 V
0 1146 R
0 -63 V
1815 251 M
0 63 V
0 1146 R
0 -63 V
1989 251 M
0 63 V
0 1146 R
0 -63 V
2162 251 M
0 63 V
0 1146 R
0 -63 V
2336 251 M
0 63 V
0 1146 R
0 -63 V
600 251 M
1736 0 V
0 1209 V
-1736 0 V
600 251 L
LT0
1007 1121 M
180 0 V
618 1460 M
17 -1 V
18 -2 V
17 -1 V
18 -2 V
17 -2 V
18 -2 V
17 -3 V
18 -3 V
17 -2 V
18 -3 V
17 -4 V
18 -3 V
17 -3 V
18 -4 V
18 -4 V
17 -4 V
18 -4 V
17 -4 V
18 -4 V
17 -5 V
18 -5 V
17 -4 V
18 -5 V
17 -5 V
18 -5 V
17 -5 V
18 -6 V
18 -5 V
17 -6 V
18 -5 V
17 -6 V
18 -6 V
17 -6 V
18 -6 V
17 -6 V
18 -7 V
17 -6 V
18 -7 V
17 -6 V
18 -7 V
17 -7 V
18 -7 V
18 -6 V
17 -8 V
18 -7 V
17 -7 V
18 -7 V
17 -8 V
18 -7 V
17 -8 V
18 -8 V
17 -7 V
18 -8 V
17 -8 V
18 -8 V
18 -8 V
17 -9 V
18 -8 V
17 -8 V
18 -9 V
17 -8 V
18 -9 V
17 -9 V
18 -9 V
17 -8 V
18 -9 V
17 -9 V
18 -10 V
17 -9 V
18 -9 V
18 -9 V
17 -10 V
18 -9 V
17 -10 V
18 -9 V
17 -10 V
18 -10 V
17 -10 V
18 -10 V
17 -10 V
18 -10 V
17 -10 V
18 -10 V
18 -11 V
17 -10 V
18 -10 V
17 -11 V
18 -11 V
17 -10 V
18 -11 V
17 -11 V
18 -11 V
17 -10 V
18 -11 V
17 -12 V
18 -11 V
17 -11 V
18 -11 V
LT1
1067 1021 D
669 1452 D
739 1441 D
808 1428 D
878 1412 D
947 1394 D
1017 1374 D
1086 1353 D
1156 1329 D
1225 1305 D
1294 1278 D
1364 1251 D
1433 1222 D
1503 1191 D
1572 1160 D
1642 1127 D
1711 1093 D
1780 1058 D
1850 1022 D
1919 985 D
1989 947 D
2058 908 D
2128 868 D
2197 827 D
2267 786 D
2336 743 D
LT2
1007 921 M
180 0 V
618 1458 M
17 -4 V
18 -4 V
17 -4 V
18 -5 V
17 -6 V
18 -5 V
17 -6 V
18 -6 V
17 -6 V
18 -7 V
17 -7 V
18 -7 V
17 -7 V
18 -7 V
18 -8 V
17 -7 V
18 -8 V
17 -8 V
18 -8 V
17 -8 V
18 -8 V
17 -9 V
18 -8 V
17 -9 V
18 -9 V
17 -9 V
18 -9 V
18 -9 V
17 -9 V
18 -10 V
17 -9 V
18 -9 V
17 -10 V
18 -10 V
17 -10 V
18 -10 V
17 -10 V
18 -10 V
17 -10 V
18 -10 V
17 -10 V
18 -11 V
18 -10 V
17 -11 V
18 -10 V
17 -11 V
18 -11 V
17 -11 V
18 -11 V
17 -11 V
18 -11 V
17 -11 V
18 -11 V
17 -11 V
18 -12 V
18 -11 V
17 -12 V
18 -11 V
17 -12 V
18 -12 V
17 -11 V
18 -12 V
17 -12 V
18 -12 V
17 -12 V
18 -12 V
17 -12 V
18 -12 V
17 -12 V
18 -13 V
18 -12 V
17 -12 V
18 -13 V
17 -12 V
18 -13 V
17 -13 V
18 -12 V
17 -13 V
18 -13 V
17 -13 V
18 -12 V
17 -13 V
18 -13 V
18 -13 V
17 -13 V
18 -14 V
17 -13 V
18 -13 V
17 -13 V
18 -14 V
17 -13 V
18 -13 V
17 -14 V
18 -13 V
17 -14 V
18 -13 V
17 -14 V
18 -14 V
LT3
1067 821 A
683 1444 A
765 1422 A
848 1394 A
931 1363 A
1013 1327 A
1096 1289 A
1179 1247 A
1261 1203 A
1344 1156 A
1427 1107 A
1509 1055 A
1592 1002 A
1675 946 A
1757 889 A
1840 830 A
1923 769 A
2005 706 A
2088 642 A
2171 576 A
2253 509 A
2336 440 A
stroke
grestore
end
showpage
}
\put(947,821){\makebox(0,0)[r]{$V\!=\!2$}}
\put(947,921){\makebox(0,0)[r]{$V\!=\!2$}}
\put(947,1021){\makebox(0,0)[r]{$V\!=\!1$}}
\put(947,1121){\makebox(0,0)[r]{$V\!=\!1$}}
\put(1468,51){\makebox(0,0){$\lambda u$}}
\put(100,855){%
\special{ps: gsave currentpoint currentpoint translate
270 rotate neg exch neg exch translate}%
\makebox(0,0)[b]{\shortstack{$\Delta E( \lambda u)$}}%
\special{ps: currentpoint grestore moveto}%
}
\put(2336,151){\makebox(0,0){0.05}}
\put(2162,151){\makebox(0,0){}}
\put(1989,151){\makebox(0,0){0.04}}
\put(1815,151){\makebox(0,0){}}
\put(1642,151){\makebox(0,0){0.03}}
\put(1468,151){\makebox(0,0){}}
\put(1294,151){\makebox(0,0){0.02}}
\put(1121,151){\makebox(0,0){}}
\put(947,151){\makebox(0,0){0.01}}
\put(774,151){\makebox(0,0){}}
\put(600,151){\makebox(0,0){0}}
\put(540,1460){\makebox(0,0)[r]{0}}
\put(540,1218){\makebox(0,0)[r]{-0.5}}
\put(540,976){\makebox(0,0)[r]{-1}}
\put(540,735){\makebox(0,0)[r]{-1.5}}
\put(540,493){\makebox(0,0)[r]{-2}}
\put(540,251){\makebox(0,0)[r]{-2.5}}
\end{picture}
  FIG. \ref{fig9}: \refstepcounter{figure} \label{fig9}
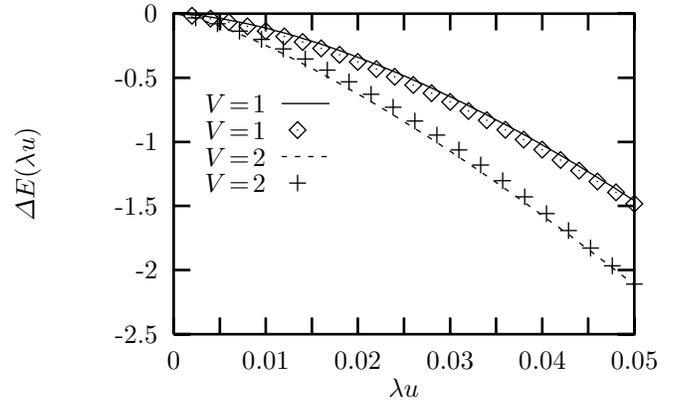
Numerical versus analytic results.  The full line corresponds to
$\Delta E(\lambda u)=-E_0[1.3 (\lambda u)^{8/5}-0.1(\lambda u)^2]$,
the dashed line to $\Delta E(\lambda u)=-E_0[0.87(\lambda u)^{4/3}-0.044(\lambda u)^2]$, where $E_0 = 2Mt/\pi$.
\\ \\
\setlength{\unitlength}{0.1bp}
\begin{picture}(2519,1511)(0,0)
\put(1468,51){\makebox(0,0){$V$}}
\put(100,855){%
\makebox(0,0)[b]{\shortstack{$u$}}%
}
\put(2191,151){\makebox(0,0){-1}}
\put(1902,151){\makebox(0,0){-1.2}}
\put(1613,151){\makebox(0,0){-1.4}}
\put(1323,151){\makebox(0,0){-1.6}}
\put(1034,151){\makebox(0,0){-1.8}}
\put(745,151){\makebox(0,0){-2}}
\put(540,1350){\makebox(0,0)[r]{1}}
\put(540,1130){\makebox(0,0)[r]{0.8}}
\put(540,910){\makebox(0,0)[r]{0.6}}
\put(540,691){\makebox(0,0)[r]{0.4}}
\put(540,471){\makebox(0,0)[r]{0.2}}
\put(540,251){\makebox(0,0)[r]{0}}
\end{picture}

 FIG. \ref{fig10}: \refstepcounter{figure} \label{fig10}
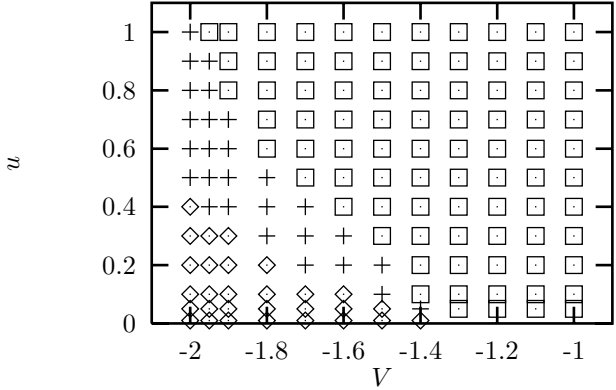
Phase diagram. The symbol $\diamond\!\negmedspace\cdot$ denotes the delocalized region 
for $\lambda_V=0$.  
The delocalized region increases when $\lambda_V=\lambda_t$, 
as indicated by the $+$. The localized region is marked by 
$\square\!\!\!\cdot\,$.
 \\  \\  
\begin{picture}(0,0)%
\epsfig{file=fig11.pstex}%
\end{picture}%
\setlength{\unitlength}{3947sp}%
\begingroup\makeatletter\ifx\SetFigFont\undefined%
\gdef\SetFigFont#1#2#3#4#5{%
  \reset@font\fontsize{#1}{#2pt}%
  \fontfamily{#3}\fontseries{#4}\fontshape{#5}%
  \selectfont}%
\fi\endgroup%
\begin{picture}(3399,2364)(1189,-2488)
\end{picture}

\\ FIG. \ref{fig11}: \refstepcounter{figure} \label{fig11}
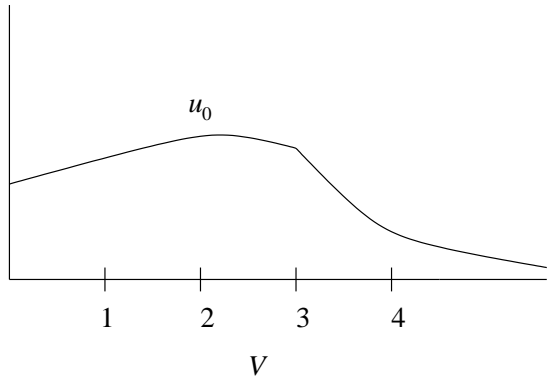
Schematic plot of  the equilibrium dimerization 
$\lambda u_0$ versus interaction at fixed $K_0$. 
 \\ 
\setlength{\unitlength}{0.1bp}
\special{!
/gnudict 40 dict def
gnudict begin
/Color false def
/Solid false def
/gnulinewidth 5.000 def
/vshift -33 def
/dl {10 mul} def
/hpt 31.5 def
/vpt 31.5 def
/M {moveto} bind def
/L {lineto} bind def
/R {rmoveto} bind def
/V {rlineto} bind def
/vpt2 vpt 2 mul def
/hpt2 hpt 2 mul def
/Lshow { currentpoint stroke M
  0 vshift R show } def
/Rshow { currentpoint stroke M
  dup stringwidth pop neg vshift R show } def
/Cshow { currentpoint stroke M
  dup stringwidth pop -2 div vshift R show } def
/DL { Color {setrgbcolor Solid {pop []} if 0 setdash }
 {pop pop pop Solid {pop []} if 0 setdash} ifelse } def
/BL { stroke gnulinewidth 2 mul setlinewidth } def
/AL { stroke gnulinewidth 2 div setlinewidth } def
/PL { stroke gnulinewidth setlinewidth } def
/LTb { BL [] 0 0 0 DL } def
/LTa { AL [1 dl 2 dl] 0 setdash 0 0 0 setrgbcolor } def
/LT0 { PL [] 0 1 0 DL } def
/LT1 { PL [4 dl 2 dl] 0 0 1 DL } def
/LT2 { PL [2 dl 3 dl] 1 0 0 DL } def
/LT3 { PL [1 dl 1.5 dl] 1 0 1 DL } def
/LT4 { PL [5 dl 2 dl 1 dl 2 dl] 0 1 1 DL } def
/LT5 { PL [4 dl 3 dl 1 dl 3 dl] 1 1 0 DL } def
/LT6 { PL [2 dl 2 dl 2 dl 4 dl] 0 0 0 DL } def
/LT7 { PL [2 dl 2 dl 2 dl 2 dl 2 dl 4 dl] 1 0.3 0 DL } def
/LT8 { PL [2 dl 2 dl 2 dl 2 dl 2 dl 2 dl 2 dl 4 dl] 0.5 0.5 0.5 DL } def
/P { stroke [] 0 setdash
  currentlinewidth 2 div sub M
  0 currentlinewidth V stroke } def
/D { stroke [] 0 setdash 2 copy vpt add M
  hpt neg vpt neg V hpt vpt neg V
  hpt vpt V hpt neg vpt V closepath stroke
  P } def
/A { stroke [] 0 setdash vpt sub M 0 vpt2 V
  currentpoint stroke M
  hpt neg vpt neg R hpt2 0 V stroke
  } def
/B { stroke [] 0 setdash 2 copy exch hpt sub exch vpt add M
  0 vpt2 neg V hpt2 0 V 0 vpt2 V
  hpt2 neg 0 V closepath stroke
  P } def
/C { stroke [] 0 setdash exch hpt sub exch vpt add M
  hpt2 vpt2 neg V currentpoint stroke M
  hpt2 neg 0 R hpt2 vpt2 V stroke } def
/T { stroke [] 0 setdash 2 copy vpt 1.12 mul add M
  hpt neg vpt -1.62 mul V
  hpt 2 mul 0 V
  hpt neg vpt 1.62 mul V closepath stroke
  P  } def
/S { 2 copy A C} def
end
}
\begin{picture}(2519,1511)(0,0)
\special{"
gnudict begin
gsave
50 50 translate
0.100 0.100 scale
0 setgray
/Helvetica findfont 100 scalefont setfont
newpath
-500.000000 -500.000000 translate
LTa
600 251 M
1736 0 V
600 251 M
0 1209 V
LTb
600 251 M
63 0 V
1673 0 R
-63 0 V
600 393 M
63 0 V
1673 0 R
-63 0 V
600 535 M
63 0 V
1673 0 R
-63 0 V
600 678 M
63 0 V
1673 0 R
-63 0 V
600 820 M
63 0 V
1673 0 R
-63 0 V
600 962 M
63 0 V
1673 0 R
-63 0 V
600 1104 M
63 0 V
1673 0 R
-63 0 V
600 1247 M
63 0 V
1673 0 R
-63 0 V
600 1389 M
63 0 V
1673 0 R
-63 0 V
600 251 M
0 63 V
0 1146 R
0 -63 V
947 251 M
0 63 V
0 1146 R
0 -63 V
1294 251 M
0 63 V
0 1146 R
0 -63 V
1642 251 M
0 63 V
0 1146 R
0 -63 V
1989 251 M
0 63 V
0 1146 R
0 -63 V
2336 251 M
0 63 V
0 1146 R
0 -63 V
600 251 M
1736 0 V
0 1209 V
-1736 0 V
600 251 L
LT0
2041 757 D
1468 561 D
1034 409 D
890 357 D
LT1
2046 773 A
1468 596 A
1034 495 A
890 480 A
LT2
2046 792 B
1468 635 B
1034 563 B
890 557 B
LT3
602 251 M
15 6 V
18 6 V
17 7 V
17 6 V
18 7 V
17 6 V
18 7 V
17 6 V
17 7 V
18 6 V
17 6 V
17 7 V
18 6 V
17 6 V
17 7 V
18 6 V
17 6 V
17 7 V
18 6 V
17 6 V
18 6 V
17 7 V
17 6 V
18 6 V
17 6 V
17 6 V
18 7 V
17 6 V
17 6 V
18 6 V
17 6 V
18 6 V
17 6 V
17 6 V
18 7 V
17 6 V
17 6 V
18 6 V
17 6 V
17 6 V
18 6 V
17 6 V
17 6 V
18 6 V
17 6 V
18 6 V
17 6 V
17 6 V
18 6 V
17 6 V
17 6 V
18 6 V
17 6 V
17 6 V
18 6 V
17 6 V
18 6 V
17 6 V
17 6 V
18 6 V
17 6 V
17 6 V
18 5 V
17 6 V
17 6 V
18 6 V
17 6 V
17 6 V
18 6 V
17 6 V
18 6 V
17 6 V
17 6 V
18 6 V
17 6 V
17 6 V
18 6 V
17 5 V
17 6 V
18 6 V
17 6 V
18 6 V
17 6 V
17 6 V
18 6 V
17 6 V
17 6 V
18 6 V
17 6 V
17 6 V
18 6 V
17 6 V
17 6 V
18 6 V
17 6 V
18 6 V
17 6 V
17 6 V
18 6 V
17 6 V
LT4
618 471 M
17 0 V
18 0 V
17 0 V
18 0 V
17 1 V
18 0 V
17 0 V
18 0 V
17 1 V
18 0 V
17 1 V
18 1 V
17 1 V
18 2 V
18 1 V
17 2 V
18 2 V
17 2 V
18 2 V
17 3 V
18 2 V
17 3 V
18 3 V
17 3 V
18 3 V
17 3 V
18 4 V
18 3 V
17 4 V
18 3 V
17 4 V
18 4 V
17 3 V
18 4 V
17 4 V
18 4 V
17 4 V
18 5 V
17 4 V
18 4 V
17 4 V
18 5 V
18 4 V
17 5 V
18 4 V
17 5 V
18 4 V
17 5 V
18 5 V
17 4 V
18 5 V
17 5 V
18 5 V
17 4 V
18 5 V
18 5 V
17 5 V
18 5 V
17 5 V
18 5 V
17 5 V
18 5 V
17 5 V
18 5 V
17 5 V
18 5 V
17 5 V
18 5 V
17 5 V
18 5 V
18 5 V
17 5 V
18 5 V
17 6 V
18 5 V
17 5 V
18 5 V
17 5 V
18 6 V
17 5 V
18 5 V
17 5 V
18 6 V
18 5 V
17 5 V
18 6 V
17 5 V
18 5 V
17 6 V
18 5 V
17 5 V
18 6 V
17 5 V
18 5 V
17 6 V
18 5 V
17 5 V
18 6 V
LT5
618 550 M
17 0 V
18 0 V
17 0 V
18 0 V
17 0 V
18 0 V
17 0 V
18 0 V
17 0 V
18 0 V
17 0 V
18 1 V
17 0 V
18 1 V
18 0 V
17 1 V
18 1 V
17 1 V
18 2 V
17 1 V
18 1 V
17 2 V
18 2 V
17 2 V
18 2 V
17 2 V
18 2 V
18 3 V
17 2 V
18 3 V
17 2 V
18 3 V
17 3 V
18 3 V
17 3 V
18 3 V
17 3 V
18 3 V
17 4 V
18 3 V
17 4 V
18 3 V
18 4 V
17 3 V
18 4 V
17 4 V
18 4 V
17 4 V
18 4 V
17 4 V
18 4 V
17 4 V
18 4 V
17 4 V
18 4 V
18 4 V
17 5 V
18 4 V
17 4 V
18 5 V
17 4 V
18 5 V
17 4 V
18 4 V
17 5 V
18 5 V
17 4 V
18 5 V
17 4 V
18 5 V
18 5 V
17 4 V
18 5 V
17 5 V
18 5 V
17 4 V
18 5 V
17 5 V
18 5 V
17 5 V
18 4 V
17 5 V
18 5 V
18 5 V
17 5 V
18 5 V
17 5 V
18 5 V
17 5 V
18 5 V
17 5 V
18 5 V
17 5 V
18 5 V
17 5 V
18 5 V
17 5 V
18 5 V
LT6
1034 918 C
890 864 C
774 841 C
LT7
2041 1416 T
1468 1131 T
1034 930 T
890 874 T
LT8
2041 1436 S
1468 1160 S
1034 963 S
890 904 S
LT0
618 790 M
17 1 V
18 1 V
17 3 V
18 3 V
17 5 V
18 5 V
17 6 V
18 6 V
17 6 V
18 7 V
17 7 V
18 7 V
17 7 V
18 8 V
18 7 V
17 8 V
18 7 V
17 8 V
18 8 V
17 8 V
18 7 V
17 8 V
18 8 V
17 8 V
18 8 V
17 8 V
18 8 V
18 8 V
17 8 V
18 8 V
17 8 V
18 8 V
17 8 V
18 8 V
17 9 V
18 8 V
17 8 V
18 8 V
17 8 V
18 8 V
17 8 V
18 9 V
18 8 V
17 8 V
18 8 V
17 8 V
18 9 V
17 8 V
18 8 V
17 8 V
18 8 V
17 9 V
18 8 V
17 8 V
18 8 V
18 9 V
17 8 V
18 8 V
17 8 V
18 9 V
17 8 V
18 8 V
17 8 V
18 9 V
17 8 V
18 8 V
17 8 V
18 9 V
17 8 V
18 8 V
18 8 V
17 9 V
18 8 V
17 8 V
18 9 V
17 8 V
18 8 V
17 8 V
18 9 V
17 8 V
18 8 V
17 8 V
18 9 V
18 8 V
17 8 V
18 9 V
17 8 V
9 4 V
stroke
grestore
end
showpage
}
\put(1034,678){\makebox(0,0)[l]{$V\!=2$}}
\put(722,1033){\makebox(0,0)[l]{$V\!=4$}}
\put(1381,1033){\makebox(0,0)[l]{$u=0.05, \lambda_V=0 \; \; \triangle$}}
\put(1642,1176){\makebox(0,0)[l]{$u=0 \; \times$}}
\put(1381,927){\makebox(0,0)[l]{$u=0.05,  \lambda_V=\lambda_t \; \times\!\!\!\!\!\!+$}}
\put(1381,358){\makebox(0,0)[l]{$u=0.05,  \lambda_V=\lambda_t \; \square$}}
\put(1381,464){\makebox(0,0)[l]{$u=0.05, \lambda_V=0 \; +$}}
\put(1642,571){\makebox(0,0)[l]{$u=0 \; \diamond$}}
\put(1468,51){\makebox(0,0){$1/M$}}
\put(100,855){%
\special{ps: gsave currentpoint currentpoint translate
270 rotate neg exch neg exch translate}%
\makebox(0,0)[b]{\shortstack{$\Delta(u,V)$}}%
\special{ps: currentpoint grestore moveto}%
}
\put(2336,151){\makebox(0,0){0.1}}
\put(1989,151){\makebox(0,0){0.08}}
\put(1642,151){\makebox(0,0){0.06}}
\put(1294,151){\makebox(0,0){0.04}}
\put(947,151){\makebox(0,0){0.02}}
\put(600,151){\makebox(0,0){0}}
\put(540,1389){\makebox(0,0)[r]{1.6}}
\put(540,1247){\makebox(0,0)[r]{1.4}}
\put(540,1104){\makebox(0,0)[r]{1.2}}
\put(540,962){\makebox(0,0)[r]{1}}
\put(540,820){\makebox(0,0)[r]{0.8}}
\put(540,678){\makebox(0,0)[r]{0.6}}
\put(540,535){\makebox(0,0)[r]{0.4}}
\put(540,393){\makebox(0,0)[r]{0.2}}
\put(540,251){\makebox(0,0)[r]{0}}
\end{picture}
  FIG. \ref{fig12}: \refstepcounter{figure} \label{fig12}
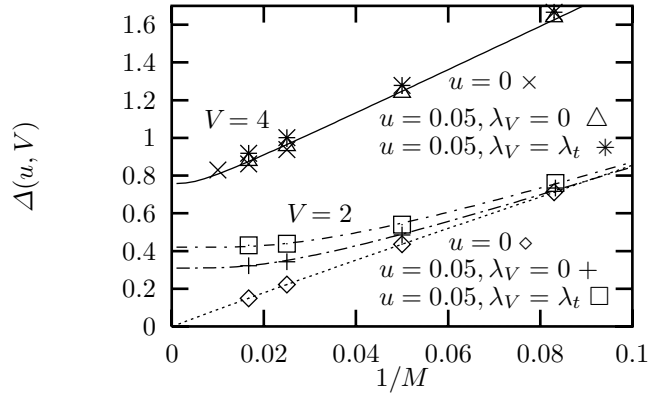
Gap versus inverse system size for various interaction strengths and dimerizations. 
The fits are extrapolations, as mentioned in the text, and are  meant
as ``guide to the eye".
\end{document}